\documentclass[usenatbib]{mnras}

\usepackage[T1]{fontenc}
\usepackage{ae,aecompl}

\usepackage{graphicx}	% Including figure files
\usepackage{amsmath}	% Advanced maths commands
\usepackage{amssymb}	% Extra maths symbols

\newcommand{\eqnumrange}[1] {\addtocounter{equation}{1}\tag{\theequation {#1}}}
 
\title[YSO variability]{Characterising the $i$-band variability of YSOs over six orders of magnitude in timescale}

\author[Darryl J. Sergison et al.]{
Darryl J. Sergison,$^{1}$
Tim Naylor,$^{1}$\thanks{E-mail: timn@astro.ex.ac.uk}
S. P. Littlefair$^{1,2}$
Cameron P. M. Bell$^{1,3}$ 
and 
\newauthor{
C. D. H. Williams$^{1}$}
\\
$^{1}$School of Physics, University of Exeter, Exeter EX4 4QL\\
$^{2}$Department of Physics \& Astronomy, University of Sheffield, Sheffield S3 7RH\\
$^{3}$Leibniz-Institut f\"ur Astrophysik Potsdam (AIP), An der Sternwarte 16, D-14482 Potsdam, Germany}

\date{Accepted 2019 December 01. Received 2019 December 01; in original form 2019 August 15}

\pubyear{2019}

\begin{document}
\label{firstpage}
\pagerange{\pageref{firstpage}--\pageref{lastpage}}
\maketitle

% Abstract of the paper
\begin{abstract}
We present an $i$-band photometric study of over 800 young stellar objects in the OB association Cep OB3b, which samples timescales from 1 minute to ten years.
Using structure functions we show that on all timescales ($\tau$) there is a monotonic decrease in variability from Class I to Class II through the transition disc (TD) systems to Class III, i.e. the more evolved systems are less variable.
The Class Is show an approximately power-law increase ($\tau^{0.8}$) in variability from timescales of a few minutes to ten years.
The Class II, TDs and Class III systems show a qualitatively different behaviour with most showing a power-law increase in variability up to a timescale corresponding to the rotational period of the star, with little additional variability beyond that timescale.
However, about a third of the Class IIs show lower overall variability, but their variability is still increasing at 10 years.
This behaviour can be explained if all Class IIs have two primary components to their variability. 
The first is an underlying roughly power-law variability spectrum, which evidence from the infrared suggests is driven by accretion rate changes.
The second component is approximately sinusoidal and results from the rotation of the star.
We suggest that the systems with dominant longer-timescale variability have a smaller rotational modulation either because they are seen at low inclinations or have more complex magnetic field geometries.

We derive a new way of calculating structure functions for large simulated datasets (the ``fast structure function''), based on fast Fourier transforms.

\end{abstract}

% Select between one and six entries from the list of approved keywords.
% Don't make up new ones.

%%%%%%%%%%%%%%%%%%%%%%%%%%%%%%%%%%%%%%%%%%%%%%%%%%

%%%%%%%%%%%%%%%%% BODY OF PAPER %%%%%%%%%%%%%%%%%%

\section{Introduction}

Accretion is a noisy process almost everywhere it occurs in astrophysics, but the resulting variability can be used to infer information about processes on length scales from the final stages of accretion onto neutron stars \citep[e.g.][]{van-der-Klis:2000aa} to the broad-line regions of active galactic nuclei \citep[e.g.][]{Peterson:2006aa}.
If young stellar objects (YSOs) are no different in this respect, there is the potential for accretion-driven variability to yield information about the final stages of accretion through the magnetosphere, and the structure of the planet-forming disc which surrounds it.
Whilst interferometry can explore the spatial scales of a large fraction of the disc size in these objects, it cannot reach the innermost disc, or the region where the disc is disrupted by the magnetic field of the young star and the material follows magnetic fields lines in its final stages of being accreted onto the star. 
Thus the amplitudes and timescales of the variability contain important clues to understanding the underlying physics of young stars.

Until recently, most work on young star variability has concentrated on searches for periodicity \citep[e.g.][]{2002A&A...396..513H,2004A&A...417..557L,2006ApJ...646..297R,Littlefair:2010aa}, typically using frequency analysis \citep[e.g.][]{1976Ap&SS..39..447L,1982ApJ...263..835S} to identify dominant periods.
This periodic variability is attributed to the presence of cool \citep[e.g.][]{1986A&A...158..149B} or hot \citep[e.g.][]{1994AJ....107.2153K} spots on the surface of the rotating star, and so the observed period is the rotation period of the star. 
Thus studying modulations at this period gives insight into the accretion structures rotating with the star, and the distribution of periods holds information about the astrophysics of the star-disc interaction.
Only one-third of YSOs, however, appear to exhibit behaviour that is broadly periodic. 
\cite{2002A&A...396..513H} found only 25-40 per cent of stars in the ONC to offer convincing periods, \cite{Littlefair:2010aa} found 25-30 per cent in Cep OB3b.

This leaves roughly half to two-thirds of the variable stars in star-forming regions without well-defined periods, and this aperiodic variability may provide a window on different physical phenomena to those associated with periodic variability. 
Aperiodic variability has often been noted and catalogued but until recently has been relatively uninterpreted within the YSO literature. 
Many stars slip between periodic and aperiodic variability \citep{1994AJ....108.1906H,2008MNRAS.391.1913R}, and many exhibit quasi-periodic lightcurve behaviour \citep[e.g.][]{2003A&A...409..169B,2014AJ....147...82C}.
For example \cite{Alencar:2010aa} find that $28\pm6$ per cent of stars monitored in NGC 2264 using the $CoRoT$ spacecraft exhibit quasi-periodic fading behaviour.
Such aperiodic or quasi-periodic variability in YSOs has been attributed to a number of mechanisms, including obscuration by circumstellar material \citep{1994AJ....108.1906H,1999A&A...345L...9C,Alencar:2010aa}, accretion shock instability \citep{2008A&A...491L..17S,2013A&A...557A..69M}, unsteady accretion \citep{1996A&A...310..143F,2009MNRAS.398..873S,2014AJ....147...83S,2014A&A...570A..82V} and instabilities within the circumstellar disc \citep{2007A&A...463.1017B, 2008ApJ...673L.171R,2011MNRAS.411..915R}.

Importantly, these phenomena can result in variability on different timescales. 
Many studies to-date have concentrated on timescales of days to a few years \citep[e.g.][]{1993A&A...272..176B,1994AJ....108.1906H, Grankin:2007aa, Rice:2012aa, 2013ApJ...773..145W}.
However there is significant evidence that we should see important variability on a much wider range of timescales. 
For example \cite{2008A&A...491L..17S} and \cite{2013A&A...557A..69M} propose that accretion shock instabilities at the surface of the star may show variability on timescales of minutes.
In contrast \cite{Contreras-Pena:2019aa} show that YSOs have outbursts on timescales of $10^4$ (Class I) to $10^5$ (Class II) years \citep[sometimes called FUor events, see][]{Reipurth:1990aa}, and \cite{2007AJ....133.2679H, Herbig:2008aa} discusses the shorter recurrence timescale EXor outbursts.
This longer-term behaviour is predicted to be driven by accretion rate variability \citep[e.g.][]{2005ApJ...633L.137V, DAngelo:2012aa},  with large surveys such as those by \cite{2017MNRAS.465.3011C} and \cite{Scholz:2013aa} providing good evidence that such long-term variability is a common phenomenon.

\subsection{Analysis Tools}

Given the above, the scientific aim of this paper is to characterise YSO variability, irrespective of whether or not that variability is periodic, and assess what this teaches us about the underlying physics.
The characterisation part of this question can be summarised as imagining we have a single epoch observation of a cluster of young stars, and asking what we could predict about their magnitudes at some future time.
To answer this question we will characterise the variability using histograms of the frequency with which a YSO is found at a particular magnitude (see Section \ref{sec:snap}), and a structure function (described in Section \ref{sec:sf_desc}), which encapsulates the degree of variability as a function of timescale.
Broadly speaking these two descriptions are orthogonal.
The structure function reflects the fact that we see larger changes in magnitude if we have two observations separated by a week, than if the observations are separated by a minute.
Conversely the histograms tell us whether that variability is symmetric about a mean magnitude, or the object spends (say) longer in a faint state than a bright state.

To a degree, the choice of these analysis tools is forced upon us by the nature of our data, and our scientific aims.
With a few notable exceptions, previous datasets in this field have been well sampled for timescales of a few days, and extended over at most a few weeks, culminating in space-based data which has dense, uninterrupted coverage for over a month.
Here we have a very unevenly sampled dataset which has large periods of time which are under-sampled compared with the rotation period.
Furthermore we are expressly searching for aperiodic signals.
This has forced us to adopt structure functions, which have been only rarely used in the field, and so we also address (mainly through the appendices) how to apply these to YSOs.

Whilst we have developed the techniques for this paper, we anticipate that they will be important for at least two upcoming datasets that will also be under-sampled with respect to the rotation period.
The first of these will be the Gaia lightcurves,  which we already have a foretaste of from the Gaia alert stream \citep{Wyrzykowski:2016aa}.
The second will be LSST \citep{Ivezic:2008aa} which importantly will provide lightcurves in many colours, which have the potential to elucidate further the mechanisms driving variability \citep[see, for example][]{Froebrich:2018aa}.

\subsection{The Target and the Dataset}

The dataset we analyse in this paper is a set of Sloan \textit{i}-band (hereafter $i$) photometric observations for $\simeq$800 YSOs, which covers more than six orders of magnitude in timescale with cadences ranging from 1 minute to 10 years.
The stars in question are members of the young OB association Cep OB3b, which was chosen simply because of the availability of a dataset taken with the same instrument over many years.
Cep OB3b contains more than 3000 X-ray and infrared identified PMS stars \citep{2012ApJ...750..125A}. 
It is similar in membership and overall size to the ONC, however Cep OB3b appears older and more evolved.
\cite{2013MNRAS.434..806B} assigns Cep OB3b an age of $\simeq$6 Myr (similar to $\sigma$ Ori and IC 348), though a reassessment of that age is required given the Gaia DR2 distance.
\cite{2012ApJ...750..125A} measure an average disc fraction in the region as 33$\pm2$ per cent.

\subsection{Data Tables, Figures and Symbols}
\label{sec:data_phil}

\begin{table}
\caption{Mathematical symbols used in this paper.}
\label{tab:symbols}
\begin{tabular}{ll}
\hline
Symbol & Description\\
\hline
$A_{{\rm H}68}$ & Half the magnitude range covering 68 percent of \\
 & observations. \\ 
$B$ & A fitted constant for the photon and instrument \\
& noise model. \\
$D$ & The ratio of the maximum number to median \\
& number of times a datapoint is used.\\
$F$ & Flux from a star in counts s$^{-1}$.\\
$F_i$, $F_j$ & The flux for lightcurve points $i$ and $j$. \\
$F_{\rm med}$ & The median flux in a lightcurve in counts s$^{-1}$.\\
$K_{1}$, $K_{2}$ & Fitted constants for the photon and instrument \\
& noise model. \\
$N$ & The number of datapoints in a time series. \\
$\mathcal{N}(0,\sigma)$ & A normal distribution about zero of width $\sigma$.\\
$P_k$ & The Fourier power at $f_k$. \\
$p$ & The number of datapoint pairs contributing\\
& to a measurement of the SF. \\
$p_{\rm eff} $ & The effective number of independent datapoint \\
& pairs contributing to a measurement of the SF. \\
$p_{\rm LPS}$ & As $p_{\rm eff} $ but for a nLPS at a particular magnitude. \\
$R_m$ & The autocorrelation function at time $m$. \\
$S(\tau), S_\tau$ & The value of the SF, including contributions from \\
& photon and instrument noise, at timescale $\tau$.\\
$S_{\rm brk} $ & The value of the SF at the break in \\
& power-law slope.\\
$S_{\rm LPS} $ & A value of the estimated median LPS SF.\\
$S_{\rm nett} $ & A value of the SF, after removing \\
& contributions of photon and instrument noise.\\
$Y_{ij}$ & A shortened notation for $(F_i-F_j)/F_{\rm med}$. \\
$X_k$ & The discrete Fourier transform of $x_n$.\\
$c_m$ & The $m$th element of a discrete cyclic convolution. \\
$c^\prime_m$ &The $m$th element of a discrete linear convolution. \\
$f$ & A Fourier frequency.\\
$k$ & The label for a given Fourier frequency $f$. \\
$m$ & The label for a given timescale $\tau$. \\
$n$ & The label for a given time $t$.\\
$t$ & The time of a flux measurement. \\
$t_{\rm max}$ & The time of the last flux measurement \\
& (assuming the first was at $t=0$). \\
$x_n$ & An alternative to $F_n/F_{\rm med}$ when discussing the \\
 & FSF, and an $n$-element time-series. \\
$y_n$, $z_n$ & $n$-element time-series. \\
$\alpha$ & The exponent characterising the Fourier amplitude \\
$ $ & spectrum ($f^{-\alpha}$).\\
$\beta$ & The power characterising the slope of a SF.\\
$\Delta S$ & The uncertainty in a value of $S$ or $S_{\rm nett}$.\\
$\epsilon$ & The photon and instrument noise in a \\
& measurement, in counts s$^{-1}$.\\
$\nu$ & The number of flux datapoints contributing to a \\ 
& measurement of the SF at a given $\tau$. \\
$\tau$ & The timescale for a particular value of the SF. \\
$\tau_i$, $\tau_j$ & The timescales for lightcurve points $i$ and $j$. \\
$\tau_1$, $\tau_2$ & Boundaries for SF timescale bins. \\
$\tau_{\rm brk}$ & The timescale at which there is a change in the\\
& slope of the SF.\\
$\omega$ & A frequency corresponding to a period of $\tau = {2 \pi \over{\omega}}.$ \\
\hline
\end{tabular}
\end{table}

The $\simeq$800 YSO lightcurves we analyse in this paper are a small sub-sample (chosen to be statistically appropriate for our purposes,  see Section \ref{sec:selection}) of a much larger dataset which could be used for other analyses.
We therefore made a decision early in the preparation of this paper to make all the lightcurves available.
From that position it was then logical to make the tables of data we derived from the lightcurves available as well, and so most of our figures are constructed directly from columns of those tables, often after selecting appropriate subsets.
As a result we have a tight correspondence between the data axes in our figures, the symbols used in the mathematics given in Table \ref{tab:symbols}, and the columns in our tables.
This we believe gives our work an added accessibility, and importantly makes it easy to reproduce and verify our results.

\phantom{}

\section{Observations and Data Reduction}

\subsection{Observations and photometric extraction}

We compiled time-series photometry of Cep OB3b from several observing runs carried out between 2004 and 2013.
These all used the $i$-band filter with Wide Field Camera on the Issac Newton Telescope in La Palma, and so present no problems of colour corrections between different photometric systems.
The camera has four EEV 2k$\times$4k CCDs with a pixel scale of 0.33 arcsec pixel$^{-1}$. 
The CCDs are arranged in an `L' shape providing a field of view of 34$\times$34 arcmin$^2$, but with a small square region missing from the north-west corner and approximately 1 arcmin gaps between CCDs. 
A single field was observed centred on $\alpha = 22^{\rm{h}}\,55^{\rm{m}}\,43^{\rm{s}}.3,  \delta = +62^{\rm{d}}\,40^{\rm{m}}\,13^{\rm{s}}$ J2000.0 \citep[see Fig. 1 of][] {Littlefair:2010aa}.

\begin{table}
\caption{Summary of observations made and collated as part of this study.}
\begin{center}
\begin{tabular}{lccc}
\hline
Date & Number of  & Exposure & Publication\\
& observations &  time (s)& \\
\hline
2004 Sep 21- Oct 6 & 424 & 300 & 1\\
2005 Aug 28 - Nov 1 & \phantom{0}28 & 300 & 1\\
2007 Oct 23 - 24 & \phantom{00}2 & 200 & 2\\
2007 Oct 23 - 24 & \phantom{00}1 & 500 & 2\\
2013 Nov 9 & \phantom{00}1 & 300 & - \\
2013 Nov 10 & \phantom{0}82 & \phantom{0}30 & - \\
\hline
\multicolumn{3}{l}{$^1$\cite{Littlefair:2010aa}}\\
\multicolumn{3}{l}{$^2$\cite{2013MNRAS.434..806B}}\\
\end{tabular}
\end{center}
\label{tab:cep_obs}
\end{table}

A summary of the observations is shown in Table \ref{tab:cep_obs}.
The data taken during 2004 Sep 21-Oct 6 are a time series over 16 nights with a cadence of $\simeq$8 minutes. 
Some breaks occurred due to poor weather, however observations were made for at least 2 hours on most nights.
The observations made between 2005 Aug 23 - Nov 1 typically consisted of one or two exposures per night. 
Some nights in this period were missed due to weather or telescope scheduling constraints.
Full details for these two runs and their data reduction are given by \cite{Littlefair:2010aa}.
\cite{2012MNRAS.424.3178B, 2013MNRAS.434..806B} conducted a survey of star-forming regions.
The data presented in those papers were colour-magnitude diagrams created using a range of exposures and filters.
We selected only the 200s and 500s $i$-band exposures of Cep OB3b (taken 2007 Oct 23 and 24), rejecting the shorter exposures as they would add significant inhomogeneity to the dataset.
The final set of observations were made specifically for this study and consisted of a single 300s exposure on the night of 9 November 2013, and 82 30s exposures on the following night.
The aim of the latter set of observations was to investigate variability on the shortest timescales we could, since the readout time of the CCDs (approximately 30s) limited the cadence to $\simeq$1 minute for reasonable signal-to-noise.
The night of November 10 was photometric, and the seeing for this dataset ranged from 0.9 to 1.3 arcsec with a median of 1.1 arcsec.

\subsection{Merging the datasets}
\label{sec:merge}

To combine the 2007 and 2013 data with 2004/5 dataset we extracted photometric magnitudes from the 2007 and 2013 images using star positions given in the \cite{Littlefair:2010aa} catalogue, which were translated into the co-ordinate system of each 2007 and 2013 observation using a six-coefficient model.
The data reduction followed the procedures described in \cite{2012MNRAS.424.3178B}, with the final step being the extraction of the optimal photometry as described in \cite{1998MNRAS.296..339N}.

It is normal to correct these optimally extracted magnitudes for changes in the point-spread function as a function of position in the focal plane, in a process analogous to aperture correction.
The corrections are derived by comparing optimally extracted magnitudes for bright stars to their large aperture magnitudes in the same images, and expressing these corrections as a function of position using low-order polynomials.
Having applied such corrections we compared the magnitudes of individual stars derived from single images in the 2007 and 2013 datasets with the median magnitudes from the 2004 dataset.
We found there were still slow changes in magnitude difference as a function of position in the focal plane, and sharp changes at the CCD edges.
We believe that this is caused by errors in both the flatfield correction and the profile correction.
The most likely cause of flatfield errors is scattered light in the twilight flatfields that we used for calibration, which would introduce an apparent change of sensitivity which would be a slowly changing function of position in the focal plane.
Again, this correction is normally a slow function of position in the focal plane.

We decided, therefore, to profile correct the 2007 and 2013 data not by comparing measurements of bright stars with their optimal extractions, but by comparing their optimally extracted magnitudes with their median magnitudes in the 2004 data \citep[see][for a previous use of this technique]{2013ApJS..209...28K}.
Since these median measurements have much higher signal-to-noise than large aperture measurements, they allow us to use much fainter stars for profile correction, and hence provide a much denser network of stars to be fitted by the profile-correction polynomials.
In this way we simultaneously carried out the profile correction and corrected for flatfielding errors.
We then converted all the lightcurves from corrected countrates to the Sloan AB system by applying a single zero point calculated by comparing the median 2004 fluxes with the magnitudes in the \cite{2013MNRAS.434..806B} catalogue.

The disadvantage of this profile correction scheme is that any flatfield errors in the 2004 data will become errors in the star-to-star mean magnitudes.
However, as we are interested in variability, not in star-to-star comparisons, this is a reasonable sacrifice to obtain the reduction in long-temporal-baseline drifts this procedure yields.
Finally, we applied the same profile correction procedure to all the 2004/5 data, despite the fact they had already been profile corrected, to remove any residual errors introduced by the large-aperture bright-star technique.

\begin{figure*}
\begin{center}
\includegraphics[width=2\columnwidth]{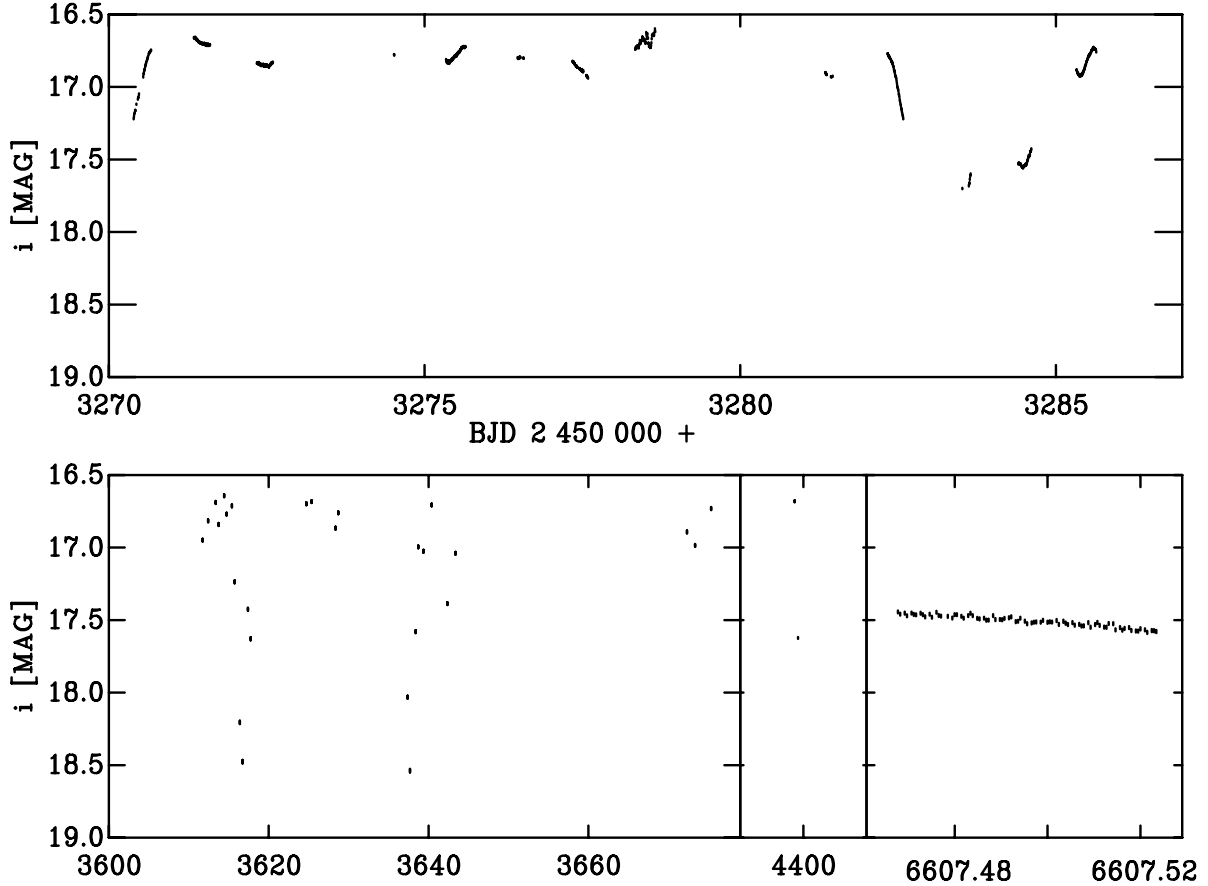}
\caption{The INT WFC $i$-band lightcurve for the Class II source 1.02 468 in Cep OB3b. 
The length of each symbol in the $y$-direction represents the 68 percent confidence limit for the measurement.
The top panel shows the data collected over 16 nights in September 2004. 
The bottom left panel is the data collected over a period of 68 nights in August to November 2005. 
The bottom middle panel is the data from October 2007, and the bottom right panel the one minute cadence data collected on a single night in November 2013.
The only datapoint not shown is the single 300s exposure on the night before the high cadence data.}
\label{fig:cep_lc}
\end{center}
\end{figure*}

The resulting set of lightcurves is given in the Electronic Table A, a description of which is given in Table \ref{tab:the_data}. 
Each point in each lightcurve consists of a barycentrically corrected mid-exposure time\footnote{We used 
\url{http://astroutils.astronomy.ohio-state.edu/time/utc2bjd.html} - see \cite{Eastman:2010aa}.}, a magnitude, an uncertainty in magnitude, a logical flag which is set true if the flux is negative (when the magnitude relates to the modulus of the flux), and quality flag which is described in detail in \cite{2003MNRAS.346.1143B}.
A set of summary statistics for each lightcurve is given in Electronic Table B, whose columns are described in Table \ref{tab:summary}.
There are just over 25 million datapoints for just over 40\,000 stars, of which 46 percent are flagged as good and have uncertainties less than 0.1 mags.
In Fig. \ref{fig:cep_lc} we show an example lightcurve which is a large amplitude Class II YSO.
The different sampling timescales for the different datasets are clearly visible, from the slow variation in the high cadence 2013 data (bottom right), through the relatively well sampled variations of the 2004 data (top panel) and the sparsely sampled 2005 data (bottom left).

\begin{table}
\caption{Column descriptions for Cep OB3b lightcurves (Electronic Table A, available from CDS at \url{http://cdsarc.u-strasbg.fr/viz-bin/qcat?II/362} or in FITS from \url{https://doi.org/10.24378/exe.2124}).}
\begin{center}
\begin{tabular}{lll}
\hline
Column Name & Units & Description \\
\hline
CCD          & & The CCD the star was \\
                  & & observed with. \\
STAR\_ID   & & Star identifier$^a$. \\
BJD           & Decimal days & Mid exposure - 2450000. \\
MAG         & Magnitudes & Sloan $i$-band magnitude. \\
UNCERT   & Magnitudes & Uncertainty in MAG.\\
FLAG         & & Quality flag$^b$.  \\
NEG\_FLUX & & Set true if flux is negative. \\
RUN\_NUM  & & INT run number. \\
\hline
\multicolumn{3}{l}{$^a$From \cite{Littlefair:2010aa}, unique only within each CCD. }\\
\multicolumn{3}{l}{$^b$See \cite{2003MNRAS.346.1143B} for the meanings of the flags.}\\
\end{tabular}
\end{center}
\label{tab:the_data}
\end{table}

\begin{table*}
\caption{Column descriptions for object properties and lightcurve summary statistics (Electronic Table B, available from CDS at \url{http://cdsarc.u-strasbg.fr/viz-bin/qcat?II/362} or in FITS from \url{https://doi.org/10.24378/exe.2124}).}
\begin{center}
\begin{tabular}{lll}
\hline
Column Name & Units & Description \\
\hline
CCD          & & CCD the star was observed with. \\
STAR\_ID   & & Star identifier$^a$. \\
i\_MAG\_CLEAN    &  Magnitudes     &  Clean $i$-band magnitude from \cite{2013MNRAS.434..806B}.\\
i\_Z\_MAG\_CLEAN   &  Magnitudes    &  Clean $i$-$Z$ colour from \cite{2013MNRAS.434..806B}. \\
NPTS & & Number of good lightcurve points$^b$. \\
A\_{\rm H}68 & Magnitudes & Half the magnitude range covering 68 percent of observations for good low-cadence datapoints. \\
MEDIAN & Magnitudes & Median magnitude for good low-cadence datapoints. \\
MEAN & Magnitudes & Mean magnitude for good low-cadence datapoints. \\
RED\_CHI & & Reduced $\chi^2$ about weighted mean of good datapoints. \\
MEAN\_UNCER & Magnitudes & Mean uncertainty for all unflagged datapoints. \\
MEDIAN\_2004 & Magnitudes & Median magnitude from 2004 data for unflagged datapoints. \\
MEDIAN\_2005 & Magnitudes & Median magnitude from 2005 data for unflagged datapoints. \\
MEDIAN\_2007 & Magnitudes & Median magnitude from 2007 data for unflagged datapoints. \\
MEDIAN\_2013 & Magnitudes & Median magnitude from 2013 data for unflagged datapoints. \\
Cl & & Class (I, II, TD, III or LPS). \\
Per & Days & Period from \cite{Littlefair:2010aa}. \\
R\_Ha\_MAG\_CLEAN & Magnitudes & Clean R$-$H$\alpha$ colour from \cite{Littlefair:2010aa}. \\
RA & Decimal degrees & J2000.0 RA from \cite{Littlefair:2010aa}. \\
Dec & Decimal degrees & J2000.0 declination from \cite{Littlefair:2010aa}. \\
\hline
\multicolumn{3}{l}{$^a$From \cite{Littlefair:2010aa}, unique only within each CCD. }\\
\multicolumn{3}{l}{$^b$I.e. unflagged datapoints with positive flux and uncertainty less than 0.2 mags. }\\
\end{tabular}
\end{center}
\label{tab:summary}
\end{table*}

Part of our purpose in presenting Electronic Table A is that we anticipate it being a useful dataset for other studies, as we have used only about two percent of the data.
We caution any potential user that it is important to examine the images\footnote{The images are available through the ING archive; \url{http://casu.ast.cam.ac.uk/casuadc/ingarch/}.}
if a result relies on a small subset of the data (especially if that subset is chosen using the data itself, e.g. looking for large amplitude variables), as the bad pixel map is not complete and the algorithm for finding non-stellar images, which should flag measurements affected by charged particle events and image blending will not always be successful. 

\subsection{Data selection and sample properties}
\label{sec:data_sel}

We chose for further analysis only those data points which have an uncertainty less than 0.2\ mags, with the data quality flag OO and that do not have negative flux.
We refer to these as ``good" datapoints. 
There about 14 million photometric datapoints which match these criteria, spread over 38\,600 stars, with a 
maximum of 540 good datapoints in any one lightcurve.
We further restrict ourselves to using just those 25\,000 lightcurves with more than 250 good datapoints and a mean uncertainty below 0.1 mags, and will refer to this as the primary sample.
Of this primary sample, 21\,000 stars have $i$-band photometry flagged OO with a signal-to-noise better than 10 in the \cite{2013MNRAS.434..806B} colour-magnitude tables, 90 percent of which are brighter than $i=22.0$.
At this magnitude the lightcurves have a mean signal to noise of around eight per point.
The bright limit of the sample is set by saturation, which results in a significant loss in the number of available datapoints in each lightcurve for stars brighter than $i=16.5$, although there are lightcurves in the primary sample (i.e. with more than 250 good datapoints) a magnitude brighter than this.

\section{ Sample Selection}
\label{sec:selection}

\begin{figure}
\includegraphics[width=\columnwidth]{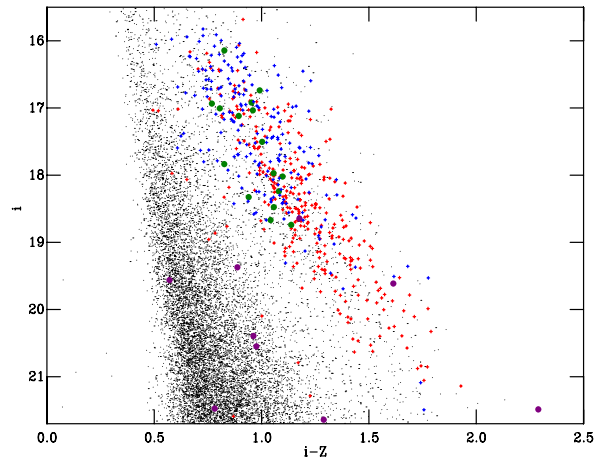}
\caption{$i$ vs $i$-$Z$ CMDs showing the classification of sources in Cep OB3b. 
The Class I (purple dots), Class II (red crosses), TD (green dots) and Class III (blue crosses) samples are overlaid on the photometry for half of all stars. 
The PMS locus is clearly visible, offset from the field stars. 
A few faint, red PMS stars lie outside the region plotted.}
\label{fig:selection}
\end{figure}

If we are to obtain a statistically robust description of YSO variability we need to begin with a well-chosen sample.
Specifically, that sample cannot be chosen using variability as part of its selection criteria.
Hence we used the classifications of \cite{2012ApJ...750..125A} which are based on \textit{Spitzer} colours and X-ray fluxes to identify stars from our sample as Class I, Class II, transition disc (TD) and Class III YSOs.
\cite{2012ApJ...750..125A} used two definitions for Class III sources.
Both first selected sources which were defined as purely photospheric on the basis of their {\it Spitzer} colours, but they then selected Class III objects from this sample either on the basis of their X-ray flux, or position in the $V$ vs $V$-$I$ colour-magnitude diagram.
We tested the efficacy of these selections, to establish which is the more appropriate for our purposes, using the fact that we expect a high fraction of Class III objects to have detectable periods.
We found that 332 out of the 1440 colour-magnitude selected Class III objects were detected as periodic in \cite{Littlefair:2010aa}.
In contrast 219 out of 499 X-ray selected objects have detected periods, suggesting the latter sample has a much lower contamination rate, in agreement with the contamination rates calculated by \cite{2012ApJ...750..125A}.
Hence we chose to use the X-ray selected objects as our sample of Class III YSOs, although this does lead to a much brighter limiting magnitude for the Class II sample compared with the Class IIIs (see Fig. \ref{fig:selection}).
The resulting classifications for each star are given in Table \ref{tab:summary}, along with summary statistics for its lightcurve.

Our primary sample (defined in Section \ref{sec:data_sel}) contains 12 Class I objects, 500 Class II stars, 21 TDs and 274 Class III stars.
% See the file Star_ids/class.num
Of these roughly a quarter of the Class II and TD stars have periods from \cite{Littlefair:2010aa}, along with about two-thirds of the Class III sources and two of the twelve Class Is.
The approximately 95 percent of the primary sample that have useful photometry in the \cite{2013MNRAS.434..806B} catalogue are shown in Fig. \ref{fig:selection}. 
Although most of the stars lie in the region of the CMD traditionally identified with PMS stars, above and to the right of the field stars, this is clearly not true of the Class I sources, where six out of the nine with sources plotted lie in the region normally associated with the contamination.
To some degree the Class I sample must be extreme objects simply because they are optically visible, and this caused us concern because, given our large sample, these could simply be mis-classifications.
However, as a class their photometric properties are different from the other classes, and we shall show below that they have very different variability.

To define the noise characteristics of our dataset, we also selected a group of local photometric standards (LPSs) by first matching against the \cite{2013MNRAS.434..806B} catalogue, and then removing those stars in the pre-main-sequence region of the $i$ vs $i-Z$ colour-magnitude diagram.
To keep the sample below 5000 stars in total (for speed of computation), we then restricted this to lightcurves with more than 536 datapoints, and finally removed the obvious outliers in a plot of mean signal-to-noise vs RMS.
Again, those stars classified as LPSs are noted as such in Table \ref{tab:summary}.

\section{Snapshot Variability}
\label{sec:snap}

As we discussed in the introduction, we divide our study of the variability into two parts, first discussing (in this section) the distribution of magnitudes an object explores as it varies, and then (in Sections \ref{sec:sf_desc} to \ref{sec:sf_disc}) discussing how those individual observations are linked in time.
This first section can be viewed as considering a series of snapshots of a cluster, where we have magnitudes for each YSO, but do not have any information as to the timing of each observation.
To characterise this aspect of YSO behaviour we must first find the distribution of the amplitudes of variability of YSOs, and then ask where within that range a star is likely to be found.

\subsection{The amplitude of variability}

The simplest measure of the amplitude of variability is the difference between the brightest and faintest magnitudes recorded for an object.
The problem with such a measure is that it will depend on how many observations are available, systematically increasing as more data are taken and the full range of variability is explored.
A simple measure which will not depend in a systematic way with the number of measurements is to use the difference between the 16th and 84th percentile points in the observed distribution of magnitudes (i.e. the range which encloses 68 percent of the data).
For Gaussian noise half of this amplitude is equal to the RMS, and so we use half the 68 per cent variability amplitude ($A_{\rm H68}$) as our metric, which should tend towards a fixed value as the number of datapoints used increases.

\begin{figure}
\includegraphics[width=\columnwidth]{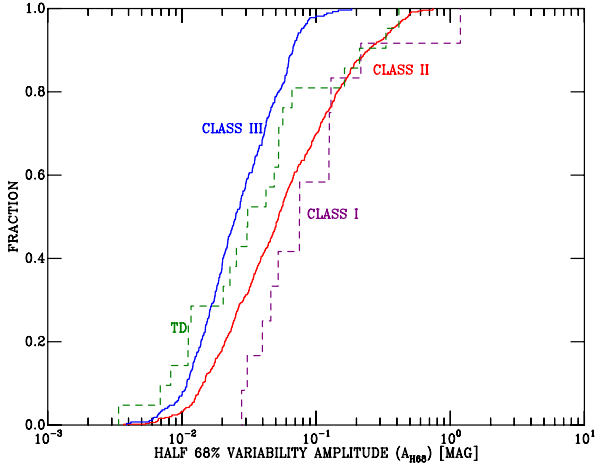}
\caption{
The cumulative distributions for half the 68 percent variability amplitude for the different classes of YSO.
Class III YSOs are shown in blue, TD systems in green, Class II in red and Class I in purple.}
\label{fig:amplitudes}
\end{figure}

\begin{figure}
\includegraphics[width=\columnwidth]{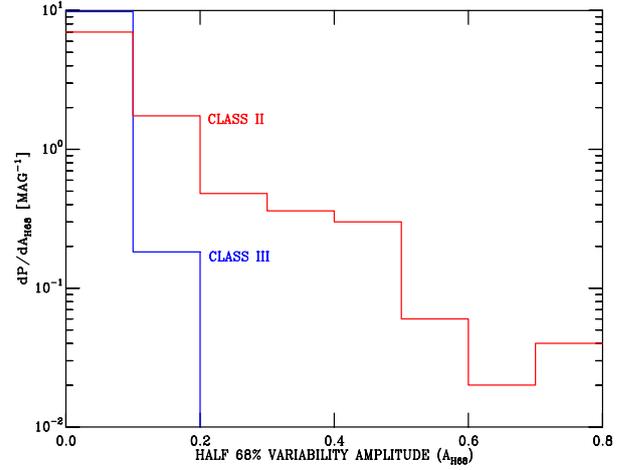}
\caption{
The differential distributions for half the 68 percent variability amplitude for Class III (blue), and Class II YSO (red).}
\label{fig:amp_diff}
\end{figure}

Using this definition of the amplitude of a star we determined the cumulative and differential distributions for each class of YSO (Fig. \ref{fig:amplitudes} and Fig. \ref{fig:amp_diff}), using just the good datapoints from the primary sample.
We also excluded the data from from the night of 2013 November 10, otherwise its high cadence means that about 10 percent of the data are at roughy the same magnitude.  

\begin{table}
\caption{Measures of overall variability as a function of YSO class.}
% Taken from Snapshot/cumulative_ii.log etc and PDF_maps/cumulative_LG10_S_BRK_III.cum etc.
\begin{center}
\begin{tabular}{lcccc}
\hline
Class & I &  II & Transition & III \\
          &   &     &  Discs \\
\hline
Median $A_{{\rm H}68}$  (mags) & 0.064 & 0.053 & 0.031 & 0.025 \\
 Median $S_{\rm brk}^{0.5}$                            & 0.13 & 0.13 & 0.10 & 0.051 \\
\hline
\end{tabular}
\end{center}
\label{tab:amps}
\end{table}

\subsection{The distribution of datapoints}
\label{sec:mag_hist}

\begin{figure}
\includegraphics[width=\columnwidth, angle=270]{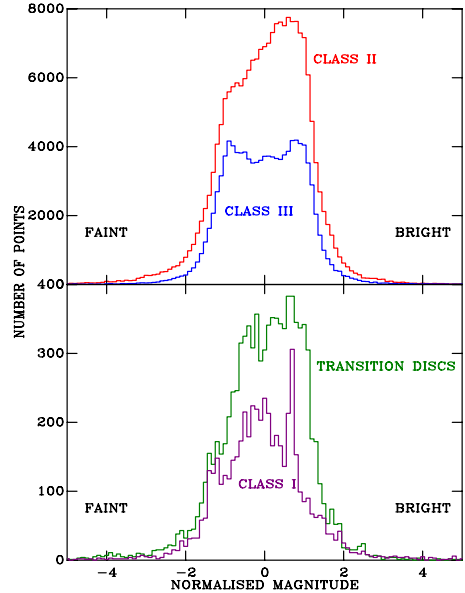}  
\caption{
The distribution of normalised magnitudes for Class II (red), Class III (blue), TD (green) and Class I (purple) YSOs.}
\label{fig:cl_hist}
\end{figure}

Whilst the amplitudes derived above represent the range of variability, they tell us nothing about how likely a star is to be found at a given point within that range.
We addressed this issue using ``normalised magnitudes'', calculated by subtracting each magnitude measurement from the mean of the good datapoints for that star, and then dividing by $A_{\rm H68}$.
We then created the distributions for each class of YSO shown in Fig. \ref{fig:cl_hist} using all the individual normalised magnitudes for each star in the class.
In doing this we had to be careful that the distribution did not become dominated by the statistical noise in each measurement, and so we selected only stars with a mean uncertainty less than 0.03 mags, except for the Class I objects, where the median uncertainty is so large (0.04 mags) that such a cut would have removed most of the sample.

\subsection{Interpreting the snapshot variability}
\label{sec:interp_snap}

It is clear from Fig. \ref{fig:amplitudes} and Fig. \ref{fig:amp_diff} that there is a decrease in variability with increasing evolutionary class, i.e. Class I objects are more variable than Class IIs, which are in turn more variable than Class IIIs.
The same trend has been found in the $JHK$ IR by \cite{Rice:2015aa}.
For the mid-range of amplitudes the different classes follow a roughly power law distribution.
Outside this range the Class II and Class III YSOs show low-amplitude tails which can be attributed to the photometric uncertainties in our data, which are typically 0.015 mags.
Presumably the same tail exists for the TD systems, but we have too few objects to resolve it.
It may also exist for the Class I YSOs as they have a median uncertainty (0.04 mags) which is much larger than the other samples, which will shift the start of the cumulative distribution to the right in Fig. \ref{fig:amplitudes}.
The Class II objects have a small tail of large amplitudes, with about 1 percent of objects showing amplitudes larger than 1 magnitude.
There is a similar tail of large amplitudes for the Class III sources, but this could be caused by contamination of the sample by Class IIs.
In summary, the distributions of the amplitudes of all the classes have very similar shapes, but scaled by a mean which decreases with evolutionary age.
The surprise here is the large amplitude of the Class I systems.
To our knowledge there is no other direct comparison of the optical variability of these stars with that of the other classes, and we shall return to their large variability after we have established their power spectrum in Section \ref{sec:sf_disc}.

\begin{figure}
\includegraphics[width=\columnwidth, angle=0]{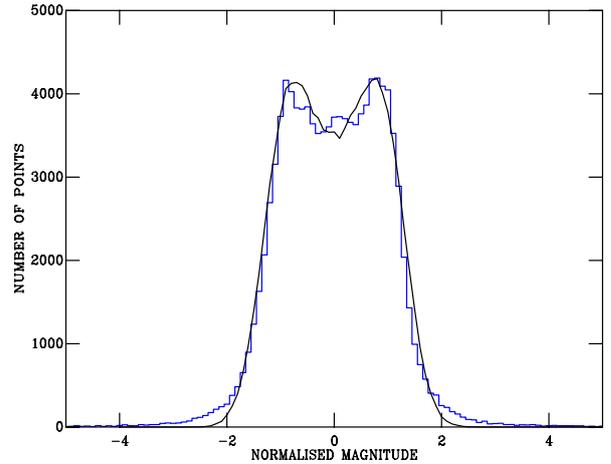}
\caption{
The distribution of normalised magnitudes for the Class III YSOs (blue histogram) with the model described in Section \ref{sec:interp_snap} overlayed as a black line.}
\label{fig:mod_hist}
\end{figure}

Given the similarity of the distribution of amplitudes, one could be tempted to assume we are looking at a single physical mechanism whose amplitude declines with time.
However, the histograms of normalised magnitudes (Fig. \ref{fig:cl_hist}) show that this is not true.
The clearest difference between classes is shown in the top panel of Fig. \ref{fig:cl_hist}, where the Class III histogram is flat-topped and symmetric, but the Class II is asymmetric.
The Class III distribution presumably comes from a modulation caused by cool spots on the surface of the star.
If we assume the modulation is sinusoidal, with an extra source of Gaussian noise we can create a model such as the one we have overlaid on the data in Fig. \ref{fig:mod_hist}.
Here the amplitude of the sine wave is 1.1 times $A_{\rm H68}$, and the Gaussian noise has an RMS of 0.4 in the same units.
As the median $A_{\rm H68}$ for the Class IIIs is about 0.025 mags, the 0.4 RMS corresponds to about 0.01 mags.
This is considerably larger than the median uncertainty for the data points in the sample, which is 0.005 mags, and so the Gaussian noise component of our model is almost certainly astrophysical in origin.
We also note that the model does not fit the extended wings of the distribution, suggesting that the noise component is non-Gaussian in the sense that the stars spend more time at extreme magnitudes than the model predicts.

Turning to the Class IIs, the sense of the asymmetry is such that the systems are spending longer in bright states than faint states, i.e. are dippers in the classification of \cite{2014AJ....147...82C}.
However, its is clear that in practice the Class IIs are a mixture of systems which dip and burst, and our data is reflecting the preponderance of dipping behaviour \citep{Findeisen:2013aa, 2014AJ....147...82C}, although it is not a strong effect, with the median only shifted by 0.1 normalised units with respect to the mean.

The Class Is are different again, lacking the flat top of the Class II sources, suggesting a concentration of magnitudes towards the mean.
However, some caution is required in over-interpreting the data since the effect of a small sample is clear: the two spikes at $-$1.3 and 0.6 mags are due to a single star (1.02 1732) which shows a deep dip. 

In summary, although the distributions of amplitudes seem to have a similar form for the different classes, albeit with a scaling, the distributions of magnitudes within those ranges are starkly different.
For the Class III objects we can understand the distribution of normalised magnitudes as noisy sine-waves, resulting from stellar activity spots on their surface.
It is unsurprising that the Class II and I distributions are different from the Class III, as we expect accretion phenomena to be the underlying mechanism powering the variability of the Class I/II YSOs.

We can summarise the results of this section by returning to our experiment of creating a set of synthetic observations of a group of YSOs knowing only their mean magnitudes.
For each YSO we would randomly select a $A_{\rm H68}$ using Fig. \ref{fig:amplitudes}, and a normalised magnitude from Fig. \ref{fig:cl_hist}.
Multiplying the two together will give the change in magnitude which should be added to the mean magnitude to yield a final simulated magnitude for each star.
We could then predict the distribution of magnitude changes for our group of YSOs between two epochs by carrying out the above procedure twice and differencing the magnitudes on a star-by-star basis.
This highlights the problem we have not addressed. 
For a real group of YSOs the distribution of changes is very different for a time interval of one minute as opposed to one year.
In the remainder of this paper we shall argue that this can be addressed using structure functions, and in doing so show they also provide clues as to what drives the shape of the magnitude distributions of the different classes.

\section{Structure Functions}
\label{sec:sf_desc}

\subsection{Introduction}

Characterising aperiodic variability is a challenge for statistical techniques.
Lomb-Scargle periodograms are successful in characterising the presence of periodic components in lightcurves. They are ineffective, however, in analysing aperiodic signals.
No standard metric analogous to the periodogram exists for characterising aperiodic signals.
Autocorrelation functions are a useful tool for finding repeating patterns in signals and may also be of use in systems where cyclical physical behaviour occurs.
However, they require that the sampling is regular and uninterrupted (though see Kreutzer et al. in prep for a possible solution to this), a situation that is very rarely achievable in astronomical datasets.
A tool that is useful for studying aperiodic signals is the structure function (SF), which has been widely used for extra-galactic work \citep[e.g.][]{1985ApJ...296...46S,1992ApJ...396..469H,2003AJ....126.1217D}.
Its use for characterising young stars has been more limited, with \cite{2017MNRAS.465.3889R} using the closely related ``pooled sigma'' for a sample of bright objects and \cite{2015ApJ...798...89F} assessing the usefulness of the related $\Delta m$-$\Delta t$ plots.

\subsection{Definition}
\label{sec:def}

The key concept of the SF is that it considers all possible pairings of the points in a lightcurve.
It is calculated in discrete logarithmically spaced timescale bins, by first taking all pairings of points in a given bin, where
\begin{equation}
\tau_{1} < \tau_{i} - \tau_{j} < \tau_{2}.
\label{eq:time_lim}
\end{equation}
and $\tau_{1}$ and $\tau_{2}$ are the lower and upper time difference limits for each bin. 
We then calculate the SF in each bin as 
\begin{equation}
S(\tau_{1}, \tau_{2}) = {\frac{1}{p(\tau_{1},\tau_{2})} \sum{{ (F_{i} - F_{j})^{2} }\over{F^2_{\rm med}}} }.
\label{eq:sf}
\end{equation}
The summation is made for all $p$($\tau_{1},\tau_{2}$) pairs of data points ($i$, $j$) with fluxes $F_{i}$ and $F_{j}$ and time separations given by Equation \ref{eq:time_lim}.
It is normal in Equation \ref{eq:sf} to divide by the mean flux of all the data points in the light curve, but we choose the median flux, $F_{\rm med}$, as it is more robust against outliers.
We label each value of the SF $S(\tau)$ where $\tau$ is the geometric mean of $\tau_1$ and $\tau_2$, but emphasise that this is a choice, driven by the logarithmic binning we will choose later.

Sometimes the SF is defined as the square root of the left-hand side of Equation \ref{eq:sf}.
This is useful in that it draws the analogy with the RMS variability, giving the values of the SF an accessible interpretation.
However, because the SF is based around two samples of the data (rather than considering the deviation of each sample from the mean), the square root of the SF is a factor of root two larger than the RMS, for a Gaussian noise signal.
In this paper we follow a middle course, and use the standard definition of the SF, but plot its square root throughout the paper, and use the same range of $S^{0.5}$ in all the plots.

\subsection{Strengths and Limitations}

The resulting SFs provide a method for assessing the frequency spectrum of the variability.
In contrast to Fourier techniques they have the advantage of being calculated in the time domain, and thus their dependence on sampling is much reduced.
Specifically gaps in the sampling have little impact on the SF, as long as a statistically meaningful number of time differences exist within a each bin. 
This makes them particularly useful in analysing data such as YSO lightcurves, where the sampling is by necessity discrete, sometimes sparse and on a wide variety of cadences.
They also have the advantage of being sensitive to all variability (including aperiodic variability) unlike a Fourier analysis which is preferentially sensitive to periodic signals.
Upper and lower limits on the timescales of physical phenomena within the system may be determined, providing the amplitude of the variability is larger than or comparable to the photometric uncertainties.
Hence SFs have been widely used in the study of active galactic nuclei, where stochastic lightcurves (similar to those often seen in YSOs) are common \citep[e.g.][]{1998ApJ...504..671K,2002MNRAS.329...76H,2008MNRAS.383.1232W,2012ApJ...753..106M}.

The main potential drawback of SFs is that they do not take any account of the uncertainties in each datapoint, which could lead to it being unclear how much of a signal is actually due to non-astrophysical noise in the data.
However, we present a solution to this problem in Appendix \ref{app:uncer} and Section \ref{sec:noise}.
Finally, it is worth noting that the requirement for a large number of datapoints in each time-domain bin leads to these bins being large (covering a range of 1.5 in time in our case).
Hence it is normal to refer to the time axis as a timescale axis; hence the symbol $\tau$ rather than $t$. 

\subsection{Simple structure functions}
\label{sec:simple}

\begin{figure}
\begin{center}
\includegraphics[width=\columnwidth]{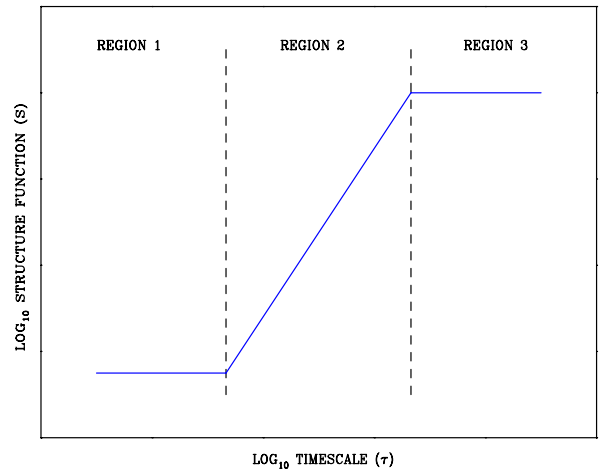}\\
\caption{A schematic SF, illustrating the main regimes (see text for detail).}
\label{fig:ideal_sf}
\end{center}
\end{figure}

Fig. \ref{fig:ideal_sf} shows an schematic SF and highlights the three main regimes.
Region 1 is at a timescale where any intrinsic variability within the lightcurve is much smaller than the measurement uncertainties for the data points. This region provides an independent estimate of the short-timescale photometric uncertainty.
Region 2 sees the SF increase with a gradient determined by the frequency spectrum of the variability exhibited by the target. 
It is possible to see plateaux in this region if the variability spectrum contains discrete low and high frequency components.
Region 3 is beyond $\tau_{\rm brk}$, the point at which the variability tends to zero. i.e. at timescales longer than $\tau_{\rm brk}$, no additional intrinsic variability, beyond that seen on shorter timescales is present.
Hence the most important conceptual difference between SFs and many other forms of power spectra is their cumulative nature.
If an object has a given variability at timescale $\tau$ we expect the SF to match or exceed that value for all timescales longer than $\tau$, though as we shall see later periodic variability can break this rule.

\subsection{Link to Fourier transforms}
\label{sec:fourier}

If region 2 is linear in log-log space, with a gradient $\beta$, it implies that the amplitude of the variability can be parameterised as a power law of the form $S \propto \tau^\beta$.
A lightcurve of random noise, i.e. one where each point is drawn randomly from the same Gaussian distribution, so there are no point-to-point correlations will give a SF where $S \propto \tau^0$.
This is the same gradient as one obtains for the Fourier power spectrum.
This leads naturally to the idea that power-law Fourier power spectra will result in a power-law form for the SF \citep[e.g.][]{Paltani:1997aa}, and that there may be a functional relationship between the two power laws \citep[e.g.][]{Voevodkin:2011aa, chapman:2005aa}. 
In fact, as we demonstrate below, this is only approximately true.
\cite{Emmanoulopoulos:2010aa} have already shown that whilst a power-law Fourier power spectrum does produce a power-law SF over much of its range, at the very longest timescales (especially the final decade), there is a tendency for the SF to flatten out.
It is concerns such as this which have led to the idea that the power law should always be measured by simulating the SF.
The problem is that good simulations require a very large number of data points, in part to overcome the noise in the simulation, but primarily to be sure one is free of sampling effects. 
For example, covering the entire time period of our the observations at the sampling of the high-cadence dataset requires approximately $5\times10^6$ datapoints, for which a typical CPU takes several hours to perform a direct evaluation of the SF.
In practice one needs to sample well beyond the timescale ``window" of the observations, and as the calculation time scales as $N^2$, suites of simulations rapidly become impractical.
However, in Appendix \ref{app:fsf} we show how it is possible to reformulate the SF using fast Fourier transforms, and create a Fast Structure Function (FSF) algorithm. 
 
 \begin{figure}
\begin{center}
\includegraphics[width=\columnwidth]{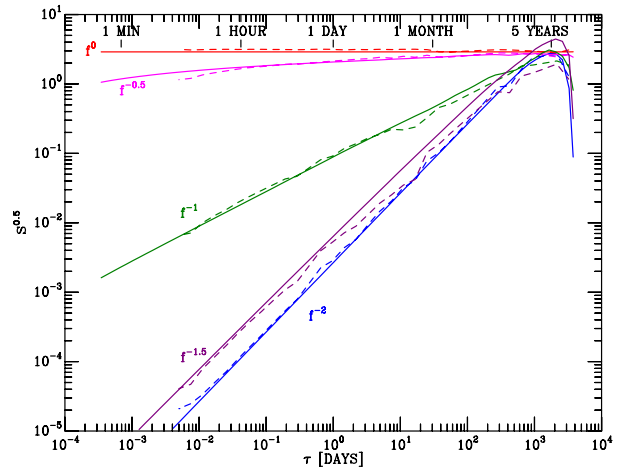}\\
\caption{Solid lines: fast structure functions for Fourier amplitudes proportional to $f^0$ (red),  $f^{-0.5}$ (fuchsia), $f^{-1}$ (green), $f^{-1.5}$ (purple) and $f^{-2}$ (blue) solid lines.
The FSFs have been normalised to similar values at 3\,000 days and placed on an arbitrary timescale.
Dashed lines: median SFs constructed from 100 realisations with the sampling of the low cadence dataset and the Fourier amplitudes as given above.
Each SF is normalised to lie close to its corresponding FSF.
}
\label{fig:power_sfs}
\end{center}
\end{figure}
 
Fig. \ref{fig:power_sfs} shows the FSFs for various Fourier amplitudes including proportional to $f^0$ (uncorrelated noise),  $f^{-0.5}$ 
\citep[flicker noise, see for example][]{1978ComAp...7..103P} and $f^{-1}$ (a random walk).
There are $2^{25}$ datapoints, which corresponds to 10.6 years with a 10\,s sampling.
As expected a Fourier power spectrum proportional to $f^0$ gives a SF of the form $S \propto \tau^0$.
A random walk gives $S \propto \tau$ over most of its length, with the roll over at long timescales seen by \cite{Emmanoulopoulos:2010aa}.
Flicker noise is more complex, with a power law close to zero, but which steepens to shorter timescales giving a mean slope of $\simeq 0.1$ over the timescales of the low-cadence dataset.
Flicker noise FSFs with fewer datapoints retain the same shape, but losing the points at shorter timescales.
Flicker noise, and our example of $f^{-1.5}$ noise show that whilst Fourier amplitudes of 0, $-1$ and $-2$ correspond to $\beta$ values of 0, 1 and 2, this simple relationship does not hold for non-integer powers.
Rather than using Gaussian noise we also tried using the PDF from the Class II YSOs (see Section \ref{sec:mag_hist} and Fig. \ref{fig:cl_hist}), but found this made little perceptible difference.

\begin{table}
\caption{Comparison of power law indices}
\begin{center}
\begin{tabular}{lcccc}
Physical  &  Fourier  & Fourier & Structure \\
Process &  Power &  Amplitude &  Function   \\
\hline
Sine function & $\delta$-fnctn &$\delta$-fnctn &$1-\cos(\omega \tau)$ \\
                     &                         &                       &$\simeq \tau^{2}$ for $\tau < {\pi\over{2\omega}}$ \\
Random walk & $f^{-2}$ & $f^{-1}$ & $\tau$ \\
Flicker noise & $f^{-1}$ & $f^{-0.5}$ & $\simeq \tau^{0.1}$ \\
Uncorrelated noise & $f^0$ & $f^0$ & $\tau^0$ \\
\end{tabular}
\end{center}
\label{tab:k_params}
\end{table}

\subsection{Structure functions with (quasi) periods}
\label{sec:period}

\begin{figure}
\begin{center}
\includegraphics[width=\columnwidth]{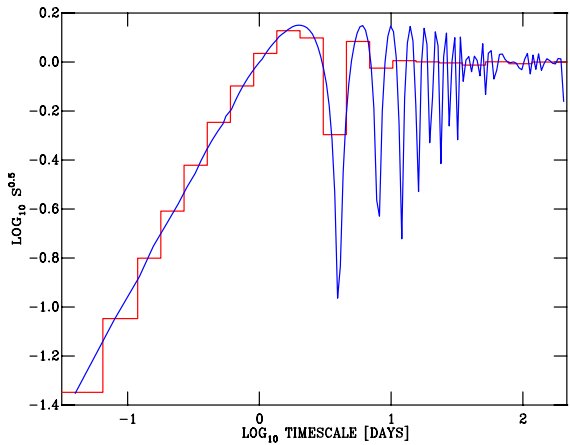}\\
\caption{The SF for a sine wave of period 4 days, where the semi-amplitude is equal to the mean.
The red line histogram uses bins where each boundary is 1.5 times the previous one, the blue curve boundaries 1.05 times the previous one.}
\label{fig:alias}
\end{center}
\end{figure}

Although the above suggests that the SF should always increase with timescale, signals containing a significant periodic component can break this rule. 
\cite{Findeisen:2015aa} shows that in the continuous case the SF of $\sin(\omega \tau)$ is 1$-$$\cos(\omega \tau)$.
Intuitively this makes sense as timescales close to the period (or an integer number of periods) will show little variability, since the lightcurve points differenced will be at similar phases.
In contrast, for timescales close to an odd integer number of half periods, the differenced points are exactly out of phase and hence show large variability. 
This results in an effect akin to aliasing, which can be seen in Fig. \ref{fig:alias}, which shows the SF for a sine wave.
However, in real cases the binning makes a significant difference.
The size of the steps chosen to sample the light curve plays an important role here, averaging out the aliasing.
Fig. \ref{fig:alias} demonstrates that the aliasing effect is much smaller when each timescale step is 1.5 times longer than the previous one (as we use later in this paper) than when the time steps are separated by a factor of 1.05.

Fig. \ref{fig:alias} also shows that a sine-wave has a nearly power-law slope corresponding  $S\propto \tau^{2}$ ($\beta$=2) for timescales shorter than about a quarter of the period.
This slope is a function of the shape of the modulation. 
Triangle functions give a similar slope, but saw-tooth or square-wave modulations give a flatter slope, because their sharp transitions introduce power at very short timescales, lifting the function at this point.

\section{Constructing the structure functions for YSOs in Cep OB3b}

\subsection{Raw structure functions}
\label{sec:raw_sf}

For each star we selected all datapoints which were unflagged and had uncertainties of less than 0.2 mags.
From these SFs were created with logarithmically spaced timescale bins in the range 0.00075 to 3686 days ($\simeq$1.0 minute to $\simeq$10 years). 
Each bin has an upper timescale limit that is 1.5 times greater than that of the previous bin.
We found that care was needed when combining the 30s exposure data from 2013 November 10 with the remaining data which is almost all 300s exposures.
This was for two reasons.
First, combining datapoints with very different signal-to-noise ratios can lead to a SF which is dominated by photon noise from the low signal-to-noise datapoints.
Second, the large number of datapoints on 2013 November 10 taken over a period of time which is much shorter than the typical variability timescale, and at a higher cadence than the remaining data can lead to a dominant variability amplitude which is not representative of the whole lightcurve, as discussed at the end of Appendix \ref{app:uncer}.
Hence, it is important that the long exposures were dealt with separately, as far as possible, from the short-exposure 2013 November 10 data. 
Hence we created two datasets. 
For for the shortest timescales ($\tau$$< $5 min) we used only the data from the night commencing 2013 November 10; we will refer to this as the high cadence dataset.
For longer timescales we binned the 30-second data from November 10 by a factor 10, so its effective exposure time matched that of the other data, and its cadence is rather slower, and then combined this with the remaining data to create what we shall refer to as the low cadence dataset.
The calculated SFs are given in Electronic Tables C and D, whose columns are described in Table \ref{tab:sf}. 

\begin{table*}
\caption{Column descriptions for tables of SFs (electronic Tables C and D, available from CDS at \url{http://cdsarc.u-strasbg.fr/viz-bin/qcat?II/362} or in FITS from \url{https://doi.org/10.24378/exe.2124}).}
\begin{center}
\begin{tabular}{llll}
\hline
Column Name & Units  & Description & Symbol \\
\hline
TAU                            & Day                   & Timescale. & $\tau$\\
S\_NETT                    &                          & SF after removal of photon and instrument noise (see  Appendix  \ref{sec:subtraction}).  & $S_{\rm nett}$ \\
DELTA\_S                  &                           & Uncertainty in S and S\_NETT as defined in  Appendix  \ref{app:uncer}. & $\Delta S$\\
S                                &                          & The SF including photon and instrument noise (see Equation \ref{eq:sf}). & $S$ \\
DELTA\_16\_S\_LPS &                           & Estimated 16th percentile SF for  an  LPS of MEDIAN\_FLUX. \\
S\_LPS                      &                           & Estimated mean SF  for  an LPS of MEDIAN\_FLUX. & $S_{\rm LPS} $ \\
DELTA\_84\_S\_LPS &                           & Estimated 84th percentile SF for  an  LPS of MEDIAN\_FLUX. \\
CCD                           &                           & The CCD the star was observed with. \\
STAR\_ID                   &                           & Star identifier$^a$. \\
MEDIAN\_FLUX         & Counts s$^{-1}$ & Median flux using conversion from magnitude for 2004 data. \\
Cl                                &                           & Class (I, II, TD, III or LPS). \\
FLAG                          &                           & Set to OU if S\_NETT $<$ DELTA\_S, otherwise OO.\\
\hline
\multicolumn{3}{l}{$^a$From \cite{Littlefair:2010aa}, unique only within each CCD. }\\
\end{tabular}
\end{center}
\label{tab:sf}
\end{table*}

\subsection{Instrument and photon noise}
\label{sec:noise}

The usual way to remove the contribution of noise induced by the instrument and of photon noise from the SF is to subtract the mean value of region 1 in Fig. \ref{fig:ideal_sf} from the SF, as any noise in this region is assumed to be non-astrophysical in origin, specifically photon noise and other sources of noise due to the instrument \citep[e.g][]{2003AJ....126.1217D}. 
The problem for our dataset is that instrument noise is not constant over all timescales.
Specifically, the long term instrumental drifts referred to in Section \ref{sec:merge} (e.g. flatfield changes), mean that the systematics in our data are larger at longer timescales.
In Appendix \ref{app:bg} we describe how we used the local standard stars to model the contributions of instrument and photon noise to the SFs and then subtracted it.
The value of the structure function after this correction we refer to as $S_{\rm nett}$. 

\subsection{The uncertainties}
\label{sec:uncer}

Given an infinite number of evenly sampled observations $S_{\rm nett}$ would be a noiseless measurement of the SF.
However, our unevenly sampled and finite timeseries means $S_{\rm nett}$ is only an estimate of the true structure function, and hence has uncertainties associated with it. 
Fig. \ref{fig:power_sfs} addresses the issue of uneven sampling by comparing well-sampled FSFs (see Appendix \ref{app:fsf}) with simulations of data sampled at the times of the low-cadence dataset \citep[see also][]{Emmanoulopoulos:2010aa}.
To mitigate the effects of the small numbers of datapoints in each timeseries, the SFs are the medians of 100 SFs.
It can be seen that our sampling has little impact on our observed SFs.

\begin{figure}
\begin{center}
\includegraphics[width=\columnwidth]{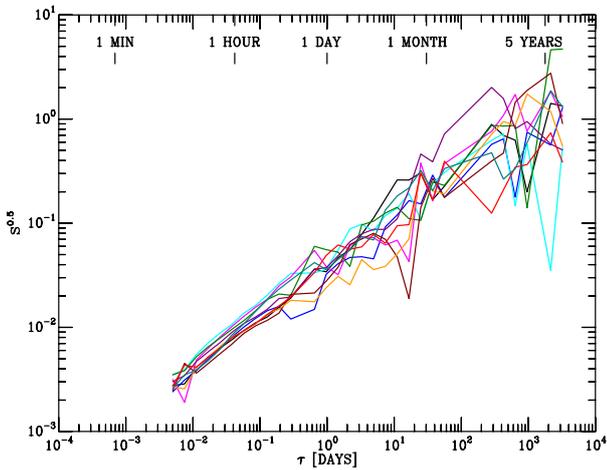}\\
\caption{Ten simulated SFs with each with a different realisation of noise whose Fourier amplitude spectrum proportional to $f^{-1}$.
The time sampling of the lightcurves is that of the low cadence dataset.}
\label{fig:walk_sfs}
\end{center}
\end{figure}

However the process we are studying is itself noisy, and so identically sampled lightcurves of the same star at different times will produce different SFs because of the limited number of samples.
This is also a function of the sampling, as different samplings may have different susceptibilities to noise.
As an example, contrast the situation where point pairs are taken from two small windows in the lightcurve separated by the timescale we are interested in, with taking the same number of pairs chosen randomly from the entire lightcurve.
If the main variability is at a timescale corresponding to the separation of the windows, the first case will give us many point pairs with very similar values, whereas the second case would truly represent the range of variability at that timescale.
Hence this issue has to be addressed for the sampling in question, and so in Fig. \ref{fig:walk_sfs} we show ten realisations of random walk noise taken on the sampling of our low-cadence dataset.
We can see that at short timescales where we have many independent samples the SF is well measured, but on longer timescales individual SFs will have uncertainties of a factor of order two.
We address this issue by simulation in Appendix \ref{app:uncer}, where we derive uncertainties ($\Delta S$) from this process, which we use in the remainder of our analysis.

\section{Interpreting the structure functions}
\label{sec:sf_disc}

\subsection{Individual structure functions}
\label{sec:individ}

\begin{figure*}
\centering
\begin{minipage}{\columnwidth}
\centering
\includegraphics[width=\columnwidth]{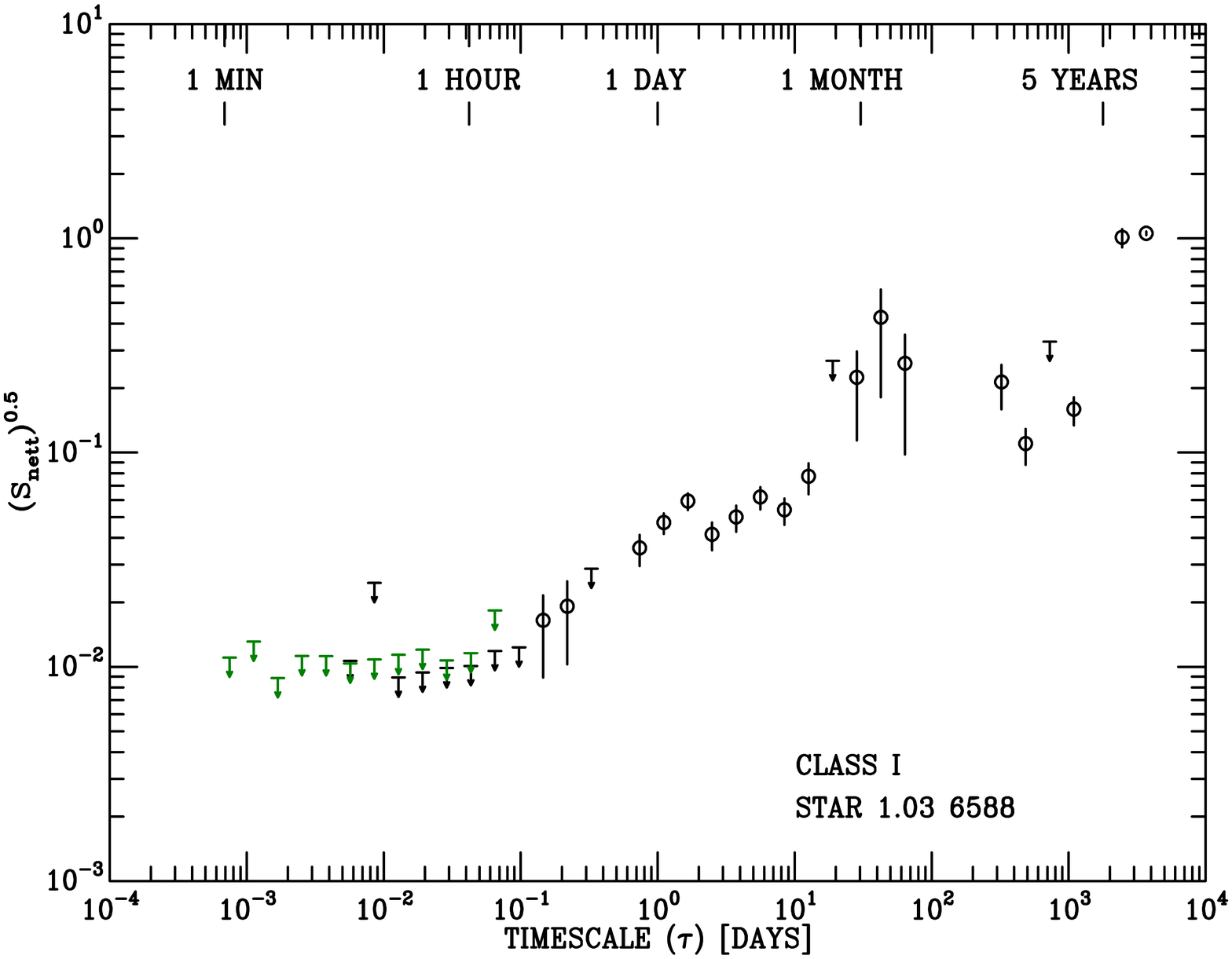}
\end{minipage}
\begin{minipage}{\columnwidth}
\centering
\includegraphics[width=\columnwidth]{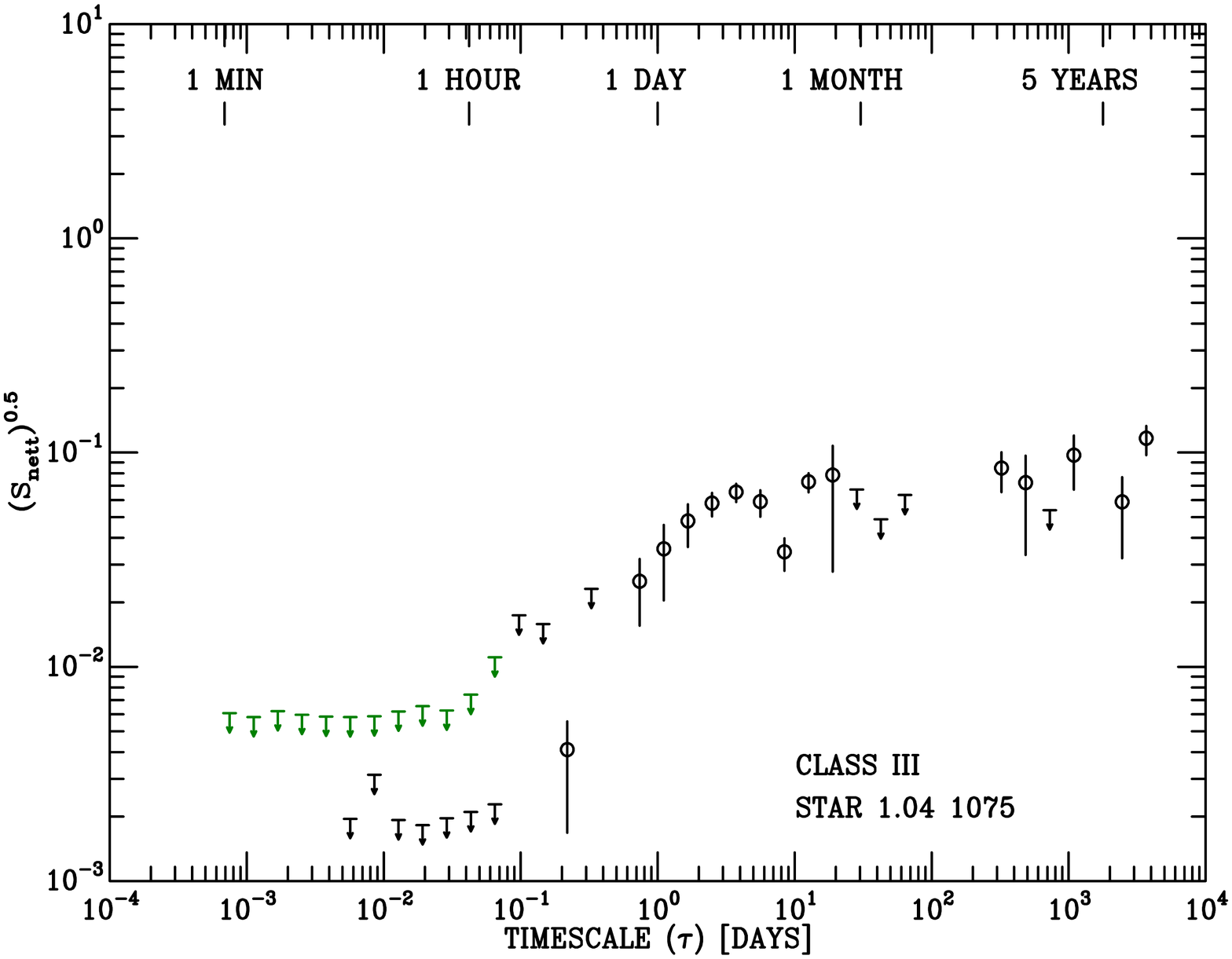}
\end{minipage}

\centering
\begin{minipage}{\columnwidth}
\centering
\includegraphics[width=\columnwidth]{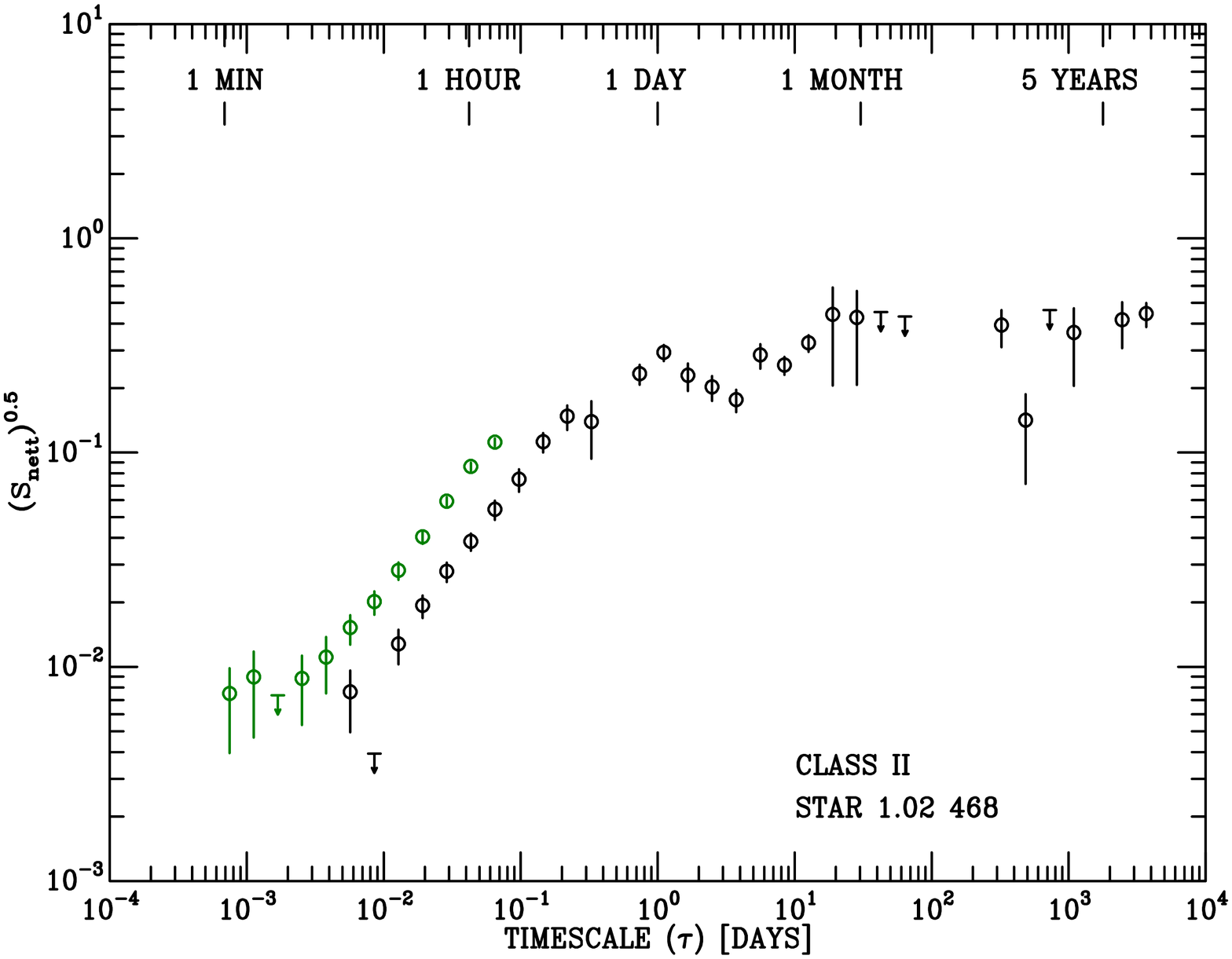}
\end{minipage}
\begin{minipage}{\columnwidth}
\centering
\includegraphics[width=\columnwidth]{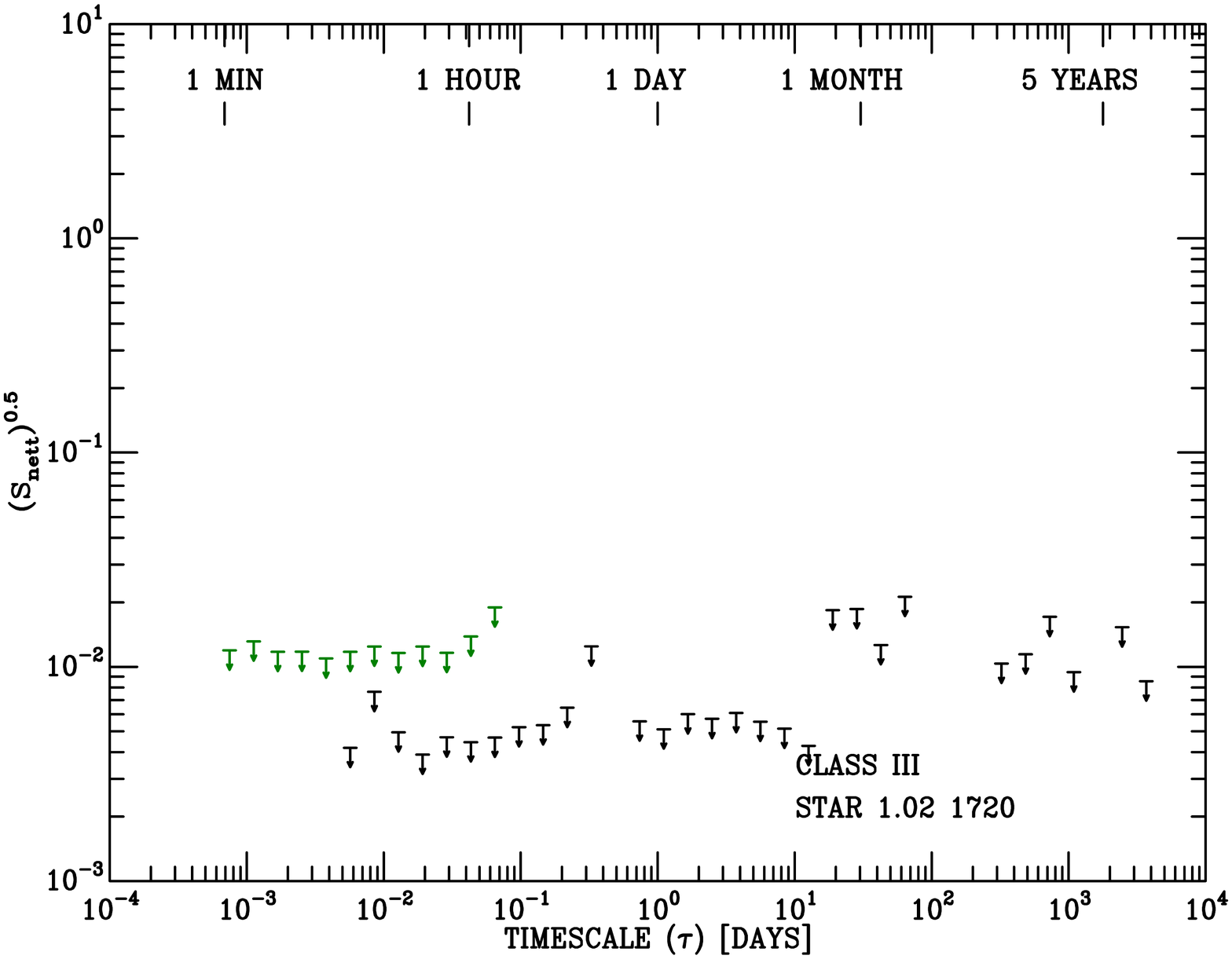}
\end{minipage}

\caption{Example SFs for YSOs. 
Circles are the SF for the target star, after the removal of photon and instrument noise, with the line showing the estimate of the uncertainty (see Section \ref{sec:uncer}).
The upper limits are the values of $\Delta S$ for datapoints where $S_{\rm nett} < \Delta S$. 
The green points are those which only use the high cadence data from 2013 November 10, whereas the black points are calculated using the low-cadence dataset.
Each panel is labelled with the identifier and classification for the star. 
%Star 1.03 6588 is a Class I YSO, 1.02 468 a Class II YSO (also shown in Figure \ref{fig:cep_lc}), and 1.04 1075 and 1.02 1720 Class III YSOs.
}
\label{fig:cep_sfs}
\end{figure*}

Examples of the calculated SFs are shown in Fig. \ref{fig:cep_sfs}.
It is immediately clear that they broadly follow the characteristics described in Section \ref{sec:simple}.
At the shortest timescales the variability is dominated by upper limits with $S^{0.5}$$\lesssim$1 per cent, which would yield a region of constant variability caused by the photon and instrument noise, had we not removed it.
Stars 1.02 468 and 1.04 1075 (Class II and III respectively) show a rise in the SF from timescales of a few minutes to around a day, after which the variability is approximately constant with timescale.
Although fitting in with this broad model, Star 1.02 468 also shows a dip centered on four days, which is almost certainly the aliasing effect referred to in Section \ref{sec:period}, since the timescales are similar to the rotation period.
Both these systems contrast with the Class I YSO (1.02 468) whose variability continues to increase even beyond timescales of 5 years.
Finally we show the Class III YSO 1.02 1720, which has a very low level variability ($A_{{\rm H}68}$=0.0086 mags) with a high mean uncertainty (0.009 mags) which means we fail to detect the variability on any timescale.
Hence the upper limits in this plot show the typical lowest level of variability we can detect.

\subsection{Median structure functions}
\label{sec:med_sf}

\begin{figure}
\begin{center}
\includegraphics[width=\columnwidth]{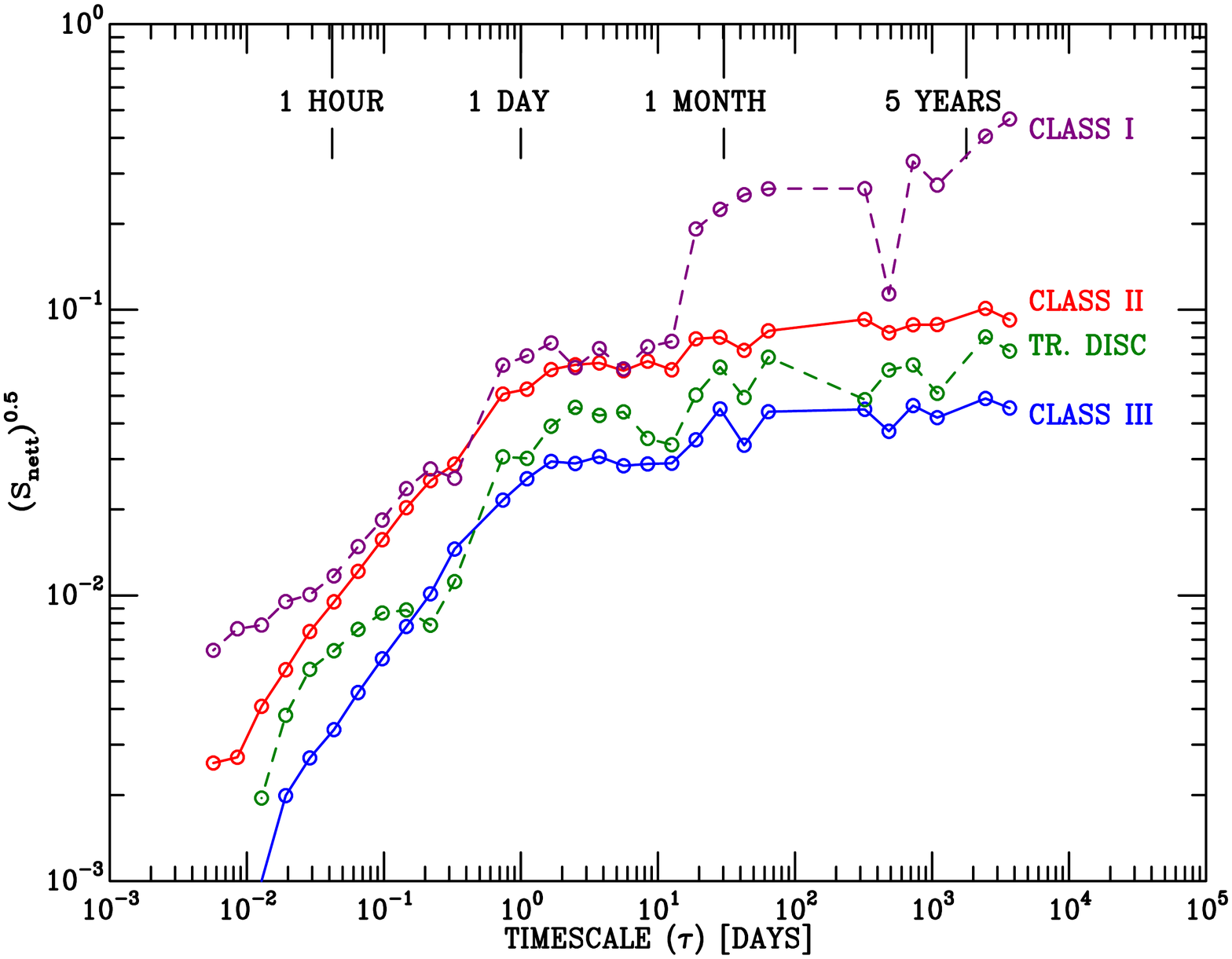}\\
\caption{Median SFs for each YSO class.}
\label{fig:median_sfs}
\end{center}
\end{figure}

To examine the differences in SF between classes we created median SFs for Classes I, II, III and TD YSOs.
To create these SFs we took the median of all the datapoints within a given timescale bin for all the stars in the clean sample.
The resulting SFs are shown in Fig. \ref{fig:median_sfs}.
For Classes II, III and the TD systems the median SFs again show the expected pattern of an increase to a break at around one day, after which the gradient is considerably smaller.
The large sample allows us to see that the rate of increase below one day is almost linear in log-log space, and is therefore power-law like.

Whilst the results from Section \ref{sec:snap} would lead us to expect a hierarchy of variability, with smaller long-term variability for more evolved stars, Fig. \ref{fig:median_sfs} shows that this hierarchy is true at all timescales.
Strikingly, the shape of the SFs are vary similar for Class II, TD and Class III YSOs, but where these systems show a break at about one day, Class I YSO variability continues to increase to the longest timescale we can sample, with a slope of $S \propto \tau^{0.8}$, which corresponds to a Fourier amplitude of roughly $f^{-0.8}$ and hence is slightly less steep than a random walk, but much steeper than ``flicker noise''.
 However, we should caution here that Figure \ref{fig:power_sfs} shows that, due to sampling effects, at timescales over 5 years the SF may underestimate the slope for noise whose amplitude spectrum is steeper than $f^{-0.5}$ (which would strengthen the result).
Conversely Figure \ref {fig:walk_sfs} shows for small numbers of objects the results at long timescales can be noisy (which would weaken the result). 

\section{Fitting the individual structure functions}
\label{subsec:fit_params}

To make further progress we need to compare the SFs with the models presented in Section \ref{sec:simple}.
We therefore fitted each YSO SF with a simple two part model where $S$ is a constant (equal to $S_{\rm brk}$) for $\tau$$>$$\tau_{\rm brk}$ and is a power law of the form $S\propto \tau^\beta$ below this. 
Hence we have
\begin{equation}
\begin{aligned}
\label{eq:pdffunc}
\log_{10} S = &\, A + \beta \log_{10} (\tau) &  & \tau < \tau_{\rm brk}\\
\log_{10} S = &\, \log_{10} (S_{\rm brk}) & & \tau > \tau_{\rm brk},
\end{aligned}
\end{equation}
where $A=\log_{10}(S_{\rm brk})-\beta \log_{10}(\tau_{\rm brk})$.
We carried out $\chi^2$ fits for all the stars in the good sample, using the uncertainties for each value of the SF derived in Appendix \ref{app:uncer}, though our use of log-log space forced us to remove SF points whose value was below zero, and we further removed any stars with fewer than 15 out of the possible 29 timescales.
We also imposed a minimum uncertainty of 0.1 dex, given that the uncertainties are probably lower limits (see Section \ref{sec:uncer}).
We grid searched the parameter ranges 
$-6 \leq \log_{10}(S_{\rm brk}) \leq 0$,
$0\leq \beta \leq3$ and 
$-2.0 \leq \log_{10}(\tau_{\rm brk}) \leq 3$; SFs where the best-fitting parameters reached these limits were removed from the analysis.

\begin{figure}
\begin{center}
\includegraphics[width=\columnwidth]{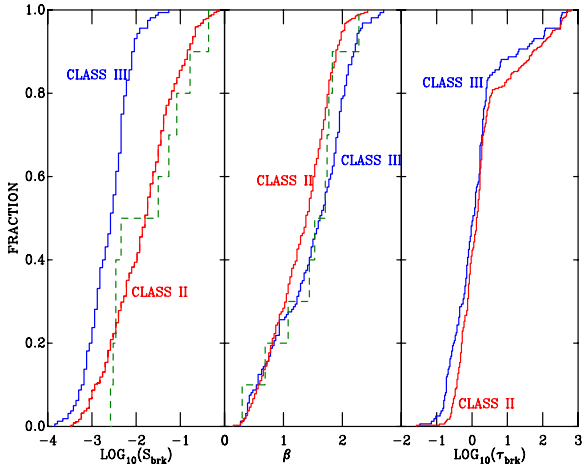}\\
\caption{
Histograms of derived parameters for YSOs in Cep OB3b. 
The left panel compares $S(\tau$) when $\tau>\tau_{\rm brk}$, for stars in the Class II (red), transition disc ( dashed green) and Class III (blue) samples. 
The center panel compares the exponent of the variability power-law ($\beta$) for the samples. 
The right panel compares $\tau_{\rm brk}$.}
\label{fig:param_fits}
\end{center}
\end{figure}

\subsection{The distributions of $S_{\rm brk}$, $\beta$ and $\tau_{\rm brk}$}

Fig. \ref{fig:param_fits} compares histograms of the derived parameters for the different YSO classes.
There are no obvious gaps in the distributions of the amplitudes ($S_{\rm brk}$, and see also Fig.\ref{fig:amp_diff}), or the timescales ($\beta$ and $\tau_{\rm brk}$), which would show as flat regions in the cumulative distributions of Fig. \ref{fig:param_fits}.
This fits with the findings of \cite{Findeisen:2013aa} who show that there is a continuum of timescale and amplitude behaviours in both bursters and faders, with \cite{Cody:2017aa} coming to a similar conclusion for bursters.
 It also fits into the broader picture drawn by \cite{Contreras-Pena:2014aa} where it is argued that the continuum of properties stretches across the canonical YSO classes. 

\subsection{Interpreting $\tau_{\rm brk}$ as the rotational period}
\label{sec:t_brk}

\begin{figure}
\begin{center}
\includegraphics[width=\columnwidth]{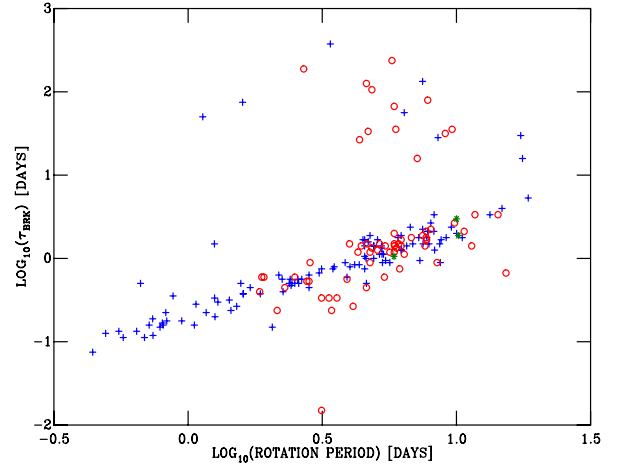}\\
\caption{$\tau_{\rm brk}$ as a function of the rotation periods for periodic stars. 
Class II (red circles), TD (green asterisks) and Class III (blue crosses) YSOs.}
\label{fig:t_brk_rot_comp}
\end{center}
\end{figure}

Fig. \ref{fig:t_brk_rot_comp} shows a comparison of $\tau_{\rm brk}$ with the rotational periods of periodic stars found by \cite{Littlefair:2010aa}. 
It is clear that $\tau_{\rm brk}$ correlates strongly with rotational period, although the value of $\tau_{\rm brk}$ is $\sim$1/4 of the period for any given star.
This matches very well to the expected break point for a sine wave sampled at our time sampling (see Section \ref{sec:period} and Fig. \ref{fig:alias}), and so we conclude that the break in the SF reflects the rotation period of the star.
Since the SFs are relatively flat beyond $\tau_{\rm brk}$, we could conclude that most of the variability of most YSOs occurs on timescales shorter than the rotational period, though there are important details hiding behind this sweeping statement.

For example, there is a group of stars at long $\tau_{\rm brk}$ where the rotational period is much shorter than $\tau_{\rm brk}$.
This means a very large fraction of their variability originates at timescales greater than their rotation periods.
We will return to the objects which lie outside the $\tau_{\rm brk}$-period relationship in Section \ref{sec:cl2_k}.

\subsection{Explaining $\beta$ for Class II YSOs}
\label{sec:cl2_k}

In discussing the origin of the slope below $\tau_{\rm brk}$, we begin by considering just the Class II systems.
Having identified $\tau_{\rm brk}$ as the rotational period of the stars, it would seem natural to associate the slope prior to the break with processes driven by the rotation.
Given that we know that many YSO lightcurves are broadly sinusoidal, we might expect the slope to correspond to that of a sine wave, i.e. $\beta$=2 (Section \ref{sec:period}).
In fact as Figs. \ref{fig:param_fits} and \ref{fig:t_brk_rot_comp} show, the gradients range from $\beta$=0.5 to only just over $\beta$=2, and so the problem becomes identifying processes which flatten the slope.
Sharp features in a strictly periodic lightcurve will decrease the slope, as we showed in Section \ref{sec:period}.
For example, whilst sine waves and triangle functions give $\beta$=2, a sawtooth or square wave will flatten the slope.
However, the phase-folded lightcurves of Class II stars are relatively time symmetric, but it is also well known that aperiodic variability occurs in YSOs, and so one or more of random walks ($\beta$=1), flicker noise ($\beta$$\simeq$0.1) and uncorrelated noise ($\beta$=0) must also contribute.

\begin{figure}
\begin{center}
\includegraphics[width=\columnwidth]{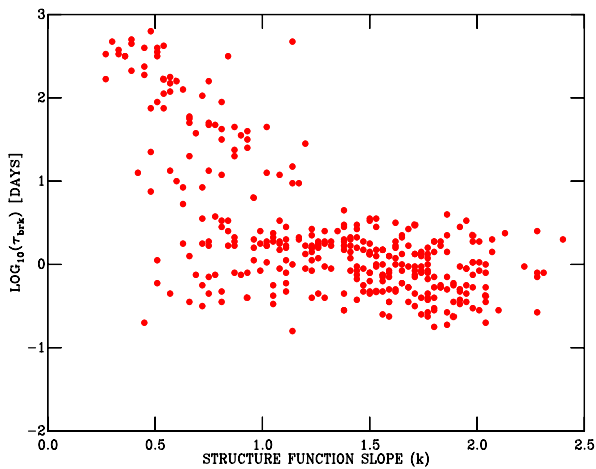}\\
\caption{$\tau_{\rm brk}$ as a function of the fitted SF slope for the Class II YSOs.}
\label{fig:k_vs_t}
\end{center}
\end{figure}

\begin{figure}
\begin{center}
\includegraphics[width=\columnwidth]{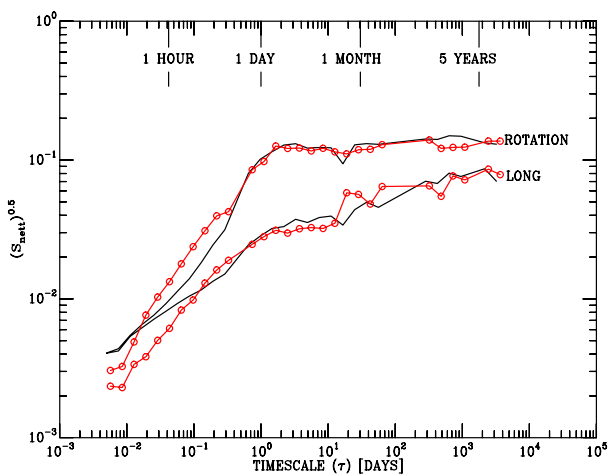}\\
\caption{
The median SFs for Class II systems which have a SF consistent with a modulation dominated by the rotation period and those which show long-timescale variability (red lines with circles).
Overlaid in black are the models described in the text.
}
\label{fig:comp_II_sfs}
\end{center}
\end{figure}

We can gain further insight by dividing our sample of Class II objects using a diagram of 
$\tau_{\rm brk}$ as a function of SF slope (Fig. \ref{fig:k_vs_t}).
Objects where $\tau_{\rm brk}$ is consistent with being a rotational period show no correlation between slope and $\tau_{\rm brk}$, whilst those with larger values of $\tau_{\rm brk}$ show an anti-correlation.
We therefore split the sample into those stars with $\tau_{\rm brk}$$<$5.4 days (or $\log_{10}\tau_{\rm brk}$$<$0.73) and those with either a longer $\tau_{\rm brk}$, or with no minimum in $\tau_{\rm brk}$ within our $\chi^2$ grid.
We show the median SFs for these groups in Fig. \ref{fig:comp_II_sfs}. 
% See the file Low/median_sfs.log for fraction in each group.
Objects below the cut-off (which consists of roughly two-thirds of the Class II stars) have a
very straightforward median SF, which at short timescales has a slope of $\tau^{1.4}$, after which there is a break at a timescale which corresponds to the rotational period, followed by a region which is entirely consistent with being flat, i.e. there is no extra variability at long timescales.
As this median SF is dominated by variability close to the rotational period, we will refer to these as the rotation group. 
For the objects where $\log_{10}\tau_{\rm brk}$$>$0.73 the SF has a slope of $\beta$$\simeq$1.0 for timescales less than a day, but then continues to grow albeit with a flatter slope to the longest timescales we can sample.
Hence we will refer to this as the long-timescale group.

What is striking is that the long-timescale group were selected for the shape of their SFs, yet they clearly have a smaller overall variability.
This suggests a model where the difference between the groups is only the amplitude of the variability at rotational timescales.
In Fig. \ref{fig:comp_II_sfs} we show that this is plausible by overlaying on the SF of the long-timescale group a model consisting of a sinusoid and a power law with Fourier amplitude proportional to $f^{-0.75}$.
The model is created by taking the median of 120 simulations each with a different period for the sinusoidal component taken from the 120 known rotation periods of the Class II primary sample.
Increasing the power of the sinusoidal component by a factor five then gives the model overlaid on the rotational group.
The models could clearly be improved at the expense of more free parameters by making the power law a weak declining function of timescale, and perhaps injecting more power at shorter timescales in the rotational component.
However, the aim with these models is to show that a single model can explain both groups, and that the mechanism which drives the largest amplitudes in Class II YSOs acts on timescales commensurate with the rotational period.
There are in addition longer-term smaller-amplitude changes, but unless the rotational variation is weak, these are swamped by the short-term variation. 

A straightforward explanation of the groups we observe might be that the rotation-dominated objects are viewed at high inclination, and so geometrical effects drive a large fraction of their variability, whereas the long-timescale group are viewed more pole-on.
Alternatively the long-timescale group may have more complex magnetic field geometries than simple dipoles.
This might lead to smaller geometric modulations in the same way as a larger number of starpots leads to a smaller rotational modulation for Class III YSOs.
Unfortunately, with the data to hand there is no way of testing these models.
For example, there is no correlation between H$\alpha$ equivalent width and inclination \cite{Appenzeller:2013aa}, despite theoretical expectations \citep{2006MNRAS.370..580K, Lima:2010aa}.
However, the inclination hypothesis could be tested using the velocity width of the H$\alpha$ line measured in high resolution data.

\subsection{Previous measurements of $\beta$ for Class II YSOs.}

Our interpretation of the SFs also fits the Fourier spectra observed by \cite{2014AJ....147...82C} (see their Fig. 17).
At high frequencies their spectra are flat, presumably corresponding to the instrumental noise in their data.
Between a tenth of a day and 10 days their spectra are proportional to $1/f$ in amplitude (equivalent to $\beta$=$1$, see Section \ref{sec:fourier}), similar to our median Class II spectra, before flattening out again at longer periods.
Interestingly the Fourier spectrum of the Herbig Ae star HD\,37806 by \cite{2010A&A...522A.113R} also shows a break at 1.5 days which may be a rotation period and again an amplitude proportional to $1/f$ below this.

In contrast \cite{2008MNRAS.391.1913R} and \cite{2016MNRAS.456.3972S} 
found a `flicker noise' spectrum in $MOST$ data of TW Hya (between 0.1 and 10 days) and RU Lup (between 0.5 and 10 days).
Given this corresponds to a $\beta$ of around $0.1$ (see Section \ref{sec:fourier}), this is much flatter than slopes we observe for our rotationally dominated group.
However RU Lup shows long-term large-amplitude variations with no sign of a rotation period \citep{2016MNRAS.456.3972S} which would place it firmly in our long-timescale group which have flatter slopes.
Likewise TW Hya is almost certainly viewed at very low inclination \citep{2016ApJ...820L..40A, Qi_2004}, and so its flat slope adds weight to our suggestion that the long-timescale group are low-inclination objects.

\subsection{Explaining $\beta$ for Class III YSOs}
\label{sec:cl3_k}

The median gradient $\beta$, for the region of the SF where $\tau < \tau_{\rm brk}$ appears to be very similar for the Class II and Class III objects. 
Considering the different physical mechanisms at play in the two populations, this might appear surprising, but in fact it shows how the SF alone does not give a full picture of the variability.
There are two key differences between the two classes.
First, if we consider the distribution of the normalised magnitudes of the points within the lightcurves (shown in Fig. \ref{fig:amplitudes}), we can see that these are very different.
The Class III objects show a double peaked distribution which, as explained in Section \ref{sec:mag_hist}, is characteristic of a sine wave with noise.
This contrasts with the single-peaked asymmetric distribution for the Class IIs. 
Second, if we split the Class III SFs in the same way we did for the Class II YSOs by the value of the break timescale we find both sub-classes tend to the same overall variability at large timescales, but the long-timescale group take longer to reach this value, again in contrast to the Class IIs.
Hence although the SFs of Class II and Class III YSOs are similar, there is other evidence which points to the differences in the mechanism producing the variability.

Fig. \ref{fig:median_sfs} shows that there is no evidence for an increase in variability on timescales longer than a month.
This agrees with the result of \cite{Grankin:2008aa}, who found very little increase in the variability of a sample of Class III YSOs on timescales of years, and it implies that although the changing morphology of the spots of the Class III sources can drive year-to-year variations in the shapes of their modulations, it does not significantly change the overall amplitude.
Given that we might expect magnetic cycles to drive changes in spot coverage, this is perhaps surprising. 

\subsection{$S_{\rm brk}$ for Class II and III YSOs}
\label{sec:sbrk}

Both \cite{Venuti:2015aa} and \cite{2017MNRAS.465.3889R} report a mean amplitude for timescales beyond the rotational ones.
From the tables in \cite{Venuti:2015aa}, the mean $r$-band RMS values for their Class II and Class III stars are 0.13 and 0.031 mags respectively, 
which is very close to our median values (0.13 and 0.051 mags respectively; see Table \ref{tab:amps}).
In contrast \cite{2017MNRAS.465.3889R} report a mean saturated variability which is much stronger at 0.22 mags (especially when it it recalled that we expect the RMS to be a factor of root two smaller than the SF; see Section \ref{sec:def}), suggesting the Class II amplitude may not be universal.
This is not surprising, we will show in Section \ref{sec:acc_vs_var} that there is a correlation between variability amplitude and accretion rate, so we would expect samples with mean ages to have have different variability, as accretion rate varies with age.

\section{The variability of Class II YSOs for $\tau$>1 year }
\label{sec:long_time}

The idea that Class II variability extends beyond the rotational timescale is not new, but is controversial.
\cite{Findeisen:2013aa} show that for a subset of sources they define as bursters and faders their variability extends out to a few hundred days.
Similarly \cite{Grankin:2007aa} find that 20 percent of their sample of Class II objects show significant variability on a timescale of years, in much the same way as we find a third of our Class II sample are in our long-timescale group, which shows variability on timescales longer than a year.

The case against such long-term variability is made by \cite{2017MNRAS.465.3889R} and \cite{Venuti:2015aa}.
The former carry out a similar analysis to ours using ``pooled sigma" which is broadly equivalent to the SF, for a sample of 39 Class II stars for timescales longer than a week.
Although they conclude that the variability on timescales of a month is no larger than that over a decade, there is evidence in their data for an increase in pooled sigma on longer timescales, though as they state, that evidence is weak.
\cite{Venuti:2015aa} present $u$- and $r$-band photometry for NGC2264, primarily taken over a fortnight, but with two epochs on years timescales. 
They conclude that there is no extra variability beyond timescales of weeks.
We think the issue here two-fold.
In the log-space in which the variability is plotted the increase in variability over the rotational timescale will always dwarf the long-term increase, even if in absolute terms they are of similar magnitude.
Fundamentally, however, we believe the difference between our work and that of \cite{2017MNRAS.465.3889R} and \cite{Venuti:2015aa} is that these two studies do not have the combination of a large sample and the long timescales required to reliably detect what is a relatively subtle effect.

In contrast to the difficulty of detecting long-term variability in the optical, long-term variability seems to be a large component of IR variability.
At $JHK$ wavelengths there is clearly increased variability as the timescale studied extends beyond a year, as demonstrated for stars in Cyg OB7
\citep{Rice:2012aa, 2013ApJ...773..145W}, the Orion Nebula Cluster \citep{Rice:2015aa} $\rho$ Oph \citep{Parks:2014aa} and the southern Galactic Plane \citep{2017MNRAS.465.3011C}.
Crucially \cite{Scholz:2012aa} detect $JHK$ variability on timescales up to 8 years for a sample drawn from many clusters.
Moving to still longer wavelengths \cite{Flaherty:2016aa} present observations at around 4$\mu$m of Chamaeleon I taken over 200 days, which show an increase in amplitude with time, and then combine their data with older IR data to show mid-IR variability extends out to a decade.
This indicates that the fraction of variability as a function of timescale is different for optical and mid-IR data.
This would be unsurprising as even at short timescales the classifications of the lightcurves are different in the different bands; few periodicities discovered in the optical are present in the mid IR, and the fluxes in the two bands are only weakly correlated \citep{2014AJ....147...82C}.
Thus it is likely that the mid-IR variability ($\simeq 4\mu$m) has a different driving mechanism from the majority of the changes in optical flux.

This behaviour is explained by the fact that the rotational modulation is believed to be primarily caused by accretion hot-spots on the surface of the star.
Hence the rotational modulation will have a lower amplitude in the infrared than the optical because the flux contrast between the hotspots and other, cooler material is smaller at longer wavelengths.
This lower amplitude then allows the longer-term variability to emerge as the dominant variation in the IR.
Interestingly, by examining the $JHK$ colours of the variability in the ONC, \cite{Rice:2015aa} make the case that the majority of the long-term IR variability is driven by accretion changes, and \cite{DAngelo:2012aa} show how the interaction of the magnetosphere with the disc might drive such accretion rate changes with the required timescale.
In combination with the results from Section \ref{sec:cl2_k} this allows us to suggest a comprehensive model for the optical and IR variability where there two components; a short-term modulation driven by geometric effects at the rotation period, and a longer-term modulation driven by changes in the accretion rate.
Due to its high temperature the rotational component can only dominate in optical observations of high inclination systems, in lower inclination systems and in the infrared accretion-rate changes will dominate the variability.

We know this is not a complete explanation of the optical lightcurves of Class II YSOs; for example dips caused by obscuration by the accretion disc must also play a role, but it may explain the majority of systems.
Furthermore there is other evidence we could collect, such as the long-term multicolour lightcurves LSST will supply. Finally, the absence of any large accretion-rate changes found by \cite{2014MNRAS.440.3444C} on timescales longer than a year from H$\alpha$ measurements (though their sample is small) argues against this model.

\section{The correlation between accretion rate and variability}
\label{sec:acc_vs_var}

\begin{figure}
\begin{center}
\includegraphics[width=\columnwidth]{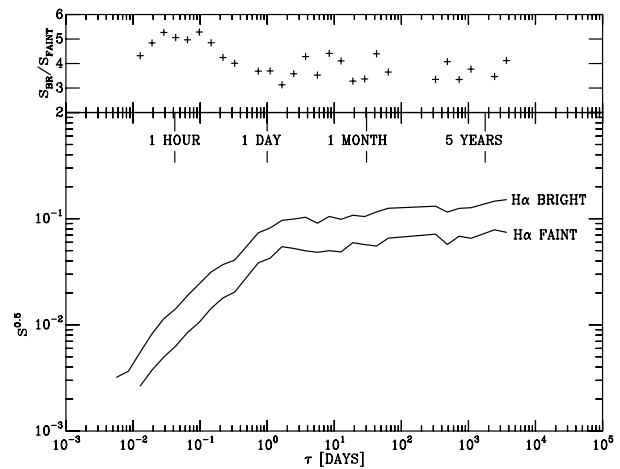}\\
\caption{
The median SFs of H$\alpha$-bright and H$\alpha$-faint stars.
The upper panel shows the ratio of the two SFs (not the ratio of $S^{0.5}$).
}
\label{fig:ha}
\end{center}
\end{figure}

To test for effects related to mean accretion rate we divided our Class II objects in the primary sample based on the $r$-$H\alpha$ colour given in \cite{Littlefair:2010aa}.
They show that the position of the pre-main-sequence in this colour is only a weak function of $V$-$I$, and so we have not performed any correction for colour.
We split the sample at $r$-$H\alpha=2.5$, which places roughly two thirds of the sample in the H$\alpha$-faint group.
We then assembled median SFs for the two samples which are presented in Fig. \ref{fig:ha}.
The most obvious result in Fig. \ref{fig:ha} is that the H$\alpha$-bright sample are about twice as variable (in terms of $S^{0.5}$, i.e RMS variability) as the faint sample.
This re-enforces the result of \cite{Cody:2017aa} who report that a sample of bursters in $\rho$ Oph and Upper Sco have stronger H$\alpha$ than non bursters.
Here, however, we can see that the increased variability is not just at the long timescales associated with bursts, indeed the ratio of the variability between the two samples is remarkably constant as a function of timescale (see the upper panel of Fig. \ref{fig:ha}) given that $S$ varies by two orders of magnitude.

Clearly, this means that higher accretion rates drive higher fractional variability, which is consistent with many theories where various instabilities increase in amplitude with increasing accretion rate.
Choosing between those theories requires understanding where in the frequency spectrum any enhanced variability occurs.
Enhanced variability at long timescales would favour theories associated with the viscous timescale of the disc, shorter timescales those associated with the rotation of the magnetosphere.
The small amplitude of the ratio in the top panel of Fig. \ref{fig:ha} argues against a mechanism acting on just one timescale.
That said, there is a case that the H$\alpha$-bright sample is more variable at short times than the H$\alpha$-faint objects.
The timescale associated with the excess variability is so short that it argues for instabilities near the shock in the accretion column, such as those modelled by \cite{2013A&A...557A..69M}.

\section{Conclusions}

Analysing the $i$-band variability of 11 Class I, 501 Class II, 21 transition disc and 274 Class III sources on timescales between 1 minute and $\simeq$10 years gives this study a unique combination of a large sample of YSOs studied over a broad range of timescales.
Variability has not been used as a selection criterion for the sample, and so the sample can be used to draw conclusions about the amplitude and prevalence of variability at the age of Cep OB3b.
The key results are as follows.
\vskip 2mm
\noindent
(i) There is a hierarchy of $i$-band optical variability between the different evolutionary classes in the sense that Class I variability is greater than Class II, which is greater than the TDs, which is in turn greater than the variability of Class III, i.e. the younger a YSO is in evolutionary terms, the more variable it is (Section \ref{sec:interp_snap} and Fig. \ref{fig:amp_diff}).
This is true at all timescales (Section \ref{sec:med_sf} and Fig. \ref{fig:median_sfs}). 
\vskip 1mm
\noindent
(ii) On timescales $\lesssim$15 minutes the $i$-band variability in the mean structure function of any class of YSO is less than $\simeq$0.2 per cent (Section \ref{sec:med_sf} and Fig. \ref{fig:median_sfs}).
\vskip 2mm
\noindent
(iii) The $i$-band variability of Class I YSOs appears to increase roughly as a power-law function of timescale up to at least ten years. 
The power-law rise ($S \propto \tau^{0.8}$) is slightly less steep than a random walk ($S \propto \tau$), but much steeper than ``flicker noise'' (Section \ref{sec:med_sf} and Fig. \ref{fig:median_sfs}).
\vskip 2mm
\noindent
(iv) On timescales between $\simeq$30 minutes and 1 day, the $i$-band variability for Class II, TD and Class III stars is entirely dominated by a power-law spectrum whose amplitude is roughly proportional to timescale to a power between 0.5 and 2.
This can be explained as a roughly sinusoidal signal at the rotational period of the star, with the addition of correlated noise (Sections \ref{sec:t_brk} and \ref{sec:cl2_k}), which probably originates from the instabilities in the accretion flow at the timescale of the inner disc \citep[see e.g.][]{2008ApJ...673L.171R}.
\vskip 2mm
\noindent
(v) Class III YSOs show no significant increase in $i$-band variability on timescales longer than a month (Section \ref{sec:cl3_k}). 
\vskip 2mm
\noindent
(vi) About two-thirds of the Class II sample show no increase in variability beyond the rotational timescale of a few days, with the remainder showing significant variability on timescales of years.
Since the sample with the longer-timescale variability also has smaller overall variability, it is clear that the amplitude of the rotational modulation is reduced.
Possible explanations for this reduction include viewing the system at relatively low inclination, or more complex field geometries (see Section \ref{sec:cl2_k}).
This reduction in turn reveals the longer-term modulation which is probably due to changes in the accretion rate (Section \ref{sec:long_time}).

\section*{Acknowledgements}

DJS was funded by a UK Science and Technology Facilities Council (STFC) studentship. 
TN is grateful to the Leverhulme Trust for a Research Project Grant without which this work would probably not have been completed.

CPMB acknowledges support from the European Research Council (ERC) under the European Union's Horizon 2020 research and innovation programme (grant agreement no. 682115).

This research is based on observations made with the Issac Newton Telescope operated on the island of La Palma by the Isaac Newton Group (ING) in the Spanish Observatorio del Roque de los Muchachos of the Instituto de Astrofisica de Canarias.
In preparing this paper we have made extensive use of TOPCAT\footnote{\url{http://www.starlink.ac.uk/topcat/}}
\citep{Taylor:2005aa}, and its script-driven equivalent STILTS\footnote{\url{http://www.starlink.ac.uk/stilts/}}
\citep{Taylor:2006aa} for exploring the data.
We thank Carlos Contreras Pe\~na for a careful reading of the final manuscript.

%%%%%%%%%%%%%%%%%%%%%%%%%%%%%%%%%%%%%%%%%%%%%%%%%%

%%%%%%%%%%%%%%%%%%%% REFERENCES %%%%%%%%%%%%%%%%%%

% The best way to enter references is to use BibTeX:

\bibliographystyle{mnras}
\bibliography{bibdesk} % if your bibtex file is called example.bib

% Alternatively you could enter them by hand, like this:
% This method is tedious and prone to error if you have lots of references
%\begin{thebibliography}{99}
%\bibitem[\protect\citeauthoryear{Author}{2012}]{Author2012}
%Author A.~N., 2013, Journal of Improbable Astronomy, 1, 1
%\bibitem[\protect\citeauthoryear{Others}{2013}]{Others2013}
%Others S., 2012, Journal of Interesting Stuff, 17, 198
%\end{thebibliography}

%%%%%%%%%%%%%%%%%%%%%%%%%%%%%%%%%%%%%%%%%%%%%%%%%%

%%%%%%%%%%%%%%%%% APPENDICES %%%%%%%%%%%%%%%%%%%%%

%\appendix

%\section{Some extra material}

%If you want to present additional material which would interrupt the flow of the main paper,
%it can be placed in an Appendix which appears after the list of references.

\appendix

\section{Fast Structure Function Simulations}
\label{app:fsf}

The simulations presented in Section \ref{sec:simple} required us to derive SFs for very large numbers of points, which we achieved by developing a Fast Structure Function (FSF) which creates SFs in a time of order
$N\log N$.
In what follows we first show how to construct time-series ($x_n$) with given noise properties using discrete Fourier transforms (DFTs), and then use these results to derive the FSF.
We also present demonstration code implementing the FSF algorithm in the supplementary material for this paper.

Since we are using DFTs, we use the variables $n$, $m$ and $k$ as integer indices labelling the observation time $t(n)$, delay time $\tau(m)$ and Fourier frequency $f(k)$ values in the usual manner, \textit{i.e.}
\begin{equation}
	t(n) = n \Delta t, \quad \tau(m) = m \Delta t, \quad f(k) = k \Delta f \eqnumrange{a--c}
\end{equation}
where
\begin{equation}
	\Delta t = t_\textrm{max}/{(N-1)}\enspace\textrm{and}\enspace\Delta f = {1}/{t_\textrm{max}} \eqnumrange{a--b}.
\end{equation}
The time-series duration $t_\textrm{max}$ must be no less than the total observation
time (9.1\,yr) and $N$ must be large enough that $\Delta t$, the time resolution, is
no greater than the shortest interval between observations (466\,s).
We assume that $N$ is an integer power of two, a choice that allows DFTs to be evaluated
in times $O(N\log N)$ using the Cooley-Tukey fast Fourier transform (FFT) algorithm \citep{cooley1965}.

\subsection{Gaussian White Noise}
\label{sec:whiteNoise} 
An \textit{N}-sample realisation of a stationary Gaussian white noise
process with zero mean and variance ${{\sigma }^{2}}$,
\begin{equation}
	{x_n}\sim \mathcal{N}\left(0,{{\sigma }^{2}}\right)\enspace \textrm{for}\enspace n=0 \dotsc (N-1)
\end{equation}
can be generated as follows. The DFT of the random vector ${x_n}$ will be
\begin{equation}
\begin{split}
	{X_k} & = \textrm{DFT}{{\left( {x_n} \right)}_k} \\
	      & = \frac{1}{N}\sum\limits_{n=0}^{N-1}{x_n}\exp \left( -i\frac{2\pi n k}{N} \right)
	      \enspace \textrm{for} \enspace k=0 \dotsc (N-1),
 \end{split}	
\end{equation}
which is a complex Gaussian random vector \citep[see, for example,][]{richards2013} with Hermitian symmetry. This can be realised directly:
\begin{subequations}
	\begin{gather}
	{X_0} \sim \mathcal{N} \left( 0,\frac{\sigma^2}{N} \right), \label{eq:XkFirst} \\
	\begin{split}
	{X_k} \sim \frac{1}{\sqrt{2}} \mathcal{N}\left( 0, \frac{\sigma^2}{N} \right)
			+\frac{i}{\sqrt{2}}\mathcal{N}\left( 0,\frac{\sigma^2}{N} \right) \\
			\enspace\textrm{for}\enspace k=1 \dotsc  (N/2)-1,
	\end{split} \\
	{X_{N/2}} \sim \mathcal{N} \left( 0,\frac{\sigma^2}{N} \right), \\
	{X_{N-k}}=X_{k}^\ast\enspace \textrm{for}\enspace k=1\dotsc (N/2)-1. \label{eq:XkLast}
	\end{gather}
\end{subequations}
The inverse discrete Fourier transform (IDFT) is then used to construct the desired white Gaussian time-series from $X_k$:
\begin{equation}\label{eq:toTimeSeries}
	\begin{split}
	x_n & =\textrm{IDFT}{\left( {X_k} \right)}_n \\
	      &=\sum\limits_{k=0}^{N-1}{{X_k}}\exp \left( +i\frac{2\pi n k}{N} \right)
	      \enspace \textrm{for} \enspace n=0 \dotsc (N-1).
	\end{split}
\end{equation}
For a given realisation, the power spectral density (PSD) is calculated directly from its definition
\begin{equation}
	P_k = X_k X_k^\ast.
\end{equation}
On average, $\left< {P_k}\right>_\textrm{white}=\sigma^2/N,$ is independent of $k$, \textit{i.e.} frequency.

\subsection{Gaussian Pink Noise}\label{sec:gaussianPinkNoise}
To simulate the astronomical observations, we need samples from a bandwidth-limited
pink-noise process with a PSD
\begin{equation}
	\left< P_k \right>_\textrm{pink} \propto
    \left\{
    	\begin{array}{ll}
    		0 & \mbox{for } k = 0, \\
    		1/f_k^{2\alpha}  & \mbox{for } 0 < k < (N-1).
    	\end{array}
    \right.
    \label{eq:pinkWeights}
\end{equation}
A sample of Gaussian noise with the desired PSD can be realised using a simple 
modification of equations \ref{eq:XkFirst}--\ref{eq:XkLast}  \citep{timmer1995}:
\begin{subequations}
	\begin{gather}
	{X_0}=0, \label{eq:pinkDC} \\
	\begin{split}
	{X_k}\sim \frac{1}{\sqrt{2}} \mathcal{N}\left(0,\left< P_k \right>_\textrm{pink}\right)+\frac{i}{\sqrt{2}}\mathcal{N}\left(0,\left< P_k \right>_\textrm{pink}\right) \\
			\enspace\textrm{for}\enspace k=1 \dotsc (N/2)-1,
	\end{split} \\
	{X_{N/2}} \sim \mathcal{N}\left(0,\left< P_k \right>_\textrm{pink}\right), \\
	{X_{N-k}}=X_k^\ast\enspace\textrm{for}\enspace k=1 \dotsc  (N/2)-1 \label{eq:pinkLast}
	\end{gather}
\end{subequations}
followed by equation~(\ref{eq:toTimeSeries}) to construct $x_n$, the required time-series.

\subsection{Non-Gaussian Pink Noise}\label{sec:generalPinkNoise}
We showed in Section \ref{sec:mag_hist} that YSO noise is non-Gaussian, and we wished to be able to simulate this for pink noise. 
To realise such a series $z_n$ we use spectral mimicry \citep{cohen1999}, which is an approximate method that was originally developed to design an experiment in microbial population dynamics. 
The steps are as follows.
\begin{enumerate}
	\item Generate $x_n$ an \textit{N}-element realisation of Gaussian noise with the desired PSD as described in section~\ref{sec:gaussianPinkNoise} above.
	\item Calculate the sort-order of the elements of $x_n$.
	This is an \textit{N}-element look-up table $n[j]$ defined such that $x_{n[j]} \le x_{n[j+1]}$ for $j=0 \dotsc N-2$.
 \item Generate $y_n$ comprising $N$ samples drawn from the desired PDF and sorted into ascending order.
 \item Calculate $z_{n[j]} = y_j$ for $j=0 \dotsc N-1$.
\end{enumerate}
The result $z_n$ is a realisation that has the desired PDF and a spectrum that approximates that of $x_n$. The execution time for this method is dominated by the sorting algorithm in step (ii), \textit{i.e.} $O(N\log N)$.

\subsection{A Fast Structure Function}
\label{sec:structFn} 
Of course it is not the $x_n$ that we are really interested in, but the SF as defined in Equation \ref{eq:sf}.
For our \textit{N}-sample realisation this becomes 
\begin{equation} \label{eq:SF}
	S_\tau = \frac{1}{(N-m)}\sum\limits_{n=0}^{N-1-m}{{{\left( {{x}_{n+m}}-{x_n} \right)}^{2}}} ,
\end{equation}
where $\tau = m \Delta t$ for $m = 0 \dotsc (N-1)$. It is not practical to evaluate equation~(\ref{eq:SF}) directly 
for large values of $N$ because this requires, see Table~\ref{tab:CPU}, a time $O(N^2)$.
Instead, we convert it into a form suitable for evaluation using FFT methods.

It is well-known that DFTs can be used to calculate cyclic convolutions \citep[][pp.~59-60]{rabiner1975}
of two \textit{N}-sample signals $x_n$ and $y_n$
\begin{equation} \label{eq:circConv}
	\begin{split}
	c_m & = \left( x_n \circledast y_n \right)_m \\
	& = \textrm{IDFT}( \textrm{DFT}(x_n) \times \textrm{DFT}(y_n) )_m \enspace\textrm{for}\enspace m = 0 \dotsc (N-1),
	\end{split}
\end{equation}
where $c_m$ also has $N$ elements. This calculation is directly applicable only to periodic signals, \textit{i.e.}
\begin{equation}
	c_m = c_{(m+N)}\quad\textrm{and}\quad x_n = x_{(n+N)}\quad\textrm{and}\quad y_n = y_{(n+N)},
\end{equation}
but the signals that interest us are aperiodic. Fortunately, this obstacle can be
overcome \citep[][pp.~61-66]{rabiner1975} by padding $x_n$ and $y_n$ with zeros, {\it e.g. }
\begin{equation}\label{eq:padding}
	x^\prime_n =
\left\{
	\begin{array}{ll}
		x_n  & \mbox{for } n  = 0\dotsc (N-1) \\
		0 & \mbox{for } n  =  N \dotsc (2N-1),
	\end{array}
\right.
\end{equation}
which facilitates calculation of the discrete linear convolution of $x_n$ and $y_n$ 
\begin{align}
	\begin{split}
	c^\prime_m (x_n,y_n)  & = \left( x_n \ast y_n \right)_m \\
	    & = \textrm{IDFT}( \textrm{DFT}(x^\prime_n) \times \textrm{DFT}(y^\prime_n) )_m \\
	     &  \qquad \qquad  \textrm{for}\enspace m = 0 \dotsc (2N-1)
	\end{split}
\end{align}
and the discrete linear autocorrelation function
\begin{align}
	\begin{split}
	R^\prime_m (x_n)  & = \frac{1}{N}\sum\limits_{n=0}^{N-1-|m|}{ {x_n } {x_{n+|m|} } } \\
	& = 2c^\prime_{|m|} (x_n,x^\ast_n) \enspace\textrm{for}\enspace m = (1-N) \dotsc (N-1).
	\end{split}
\end{align}
We use $N$ padding-zeros in equation~(\ref{eq:padding}), one more than strictly necessary, so that the length of $x^\prime_n$ remains an integer power of two, \textit{i.e.} optimal for fast calculation.

Returning our attention to the SF (eqn~\ref{eq:SF}), after
expansion of the summand it comprises three terms
\begin{equation}
	{S_\tau} = \frac{1}{N-m}\left( {S^{(a)}_\tau} + {S^{(b)}_\tau} + {S^{(c)}_\tau} \right)
\end{equation}
that can be evaluated as follows:
\begin{subequations} \label{eq:FSF}
	\begin{gather}
	{S^{(a)}_\tau} = \sum\limits_{n=0}^{N-1-m} {-2\left({x_n}{x}_{n+m}\right)} 
		= -2 N R^\prime_\tau (x_n), \\
	{S^{(b)}_\tau} =  \sum\limits_{n=0}^{N-1-m} {{x}^2_{n}}
	= \sum\limits_{n=0}^{N-1} {{x}^2_{n}} - \sum\limits_{n=0}^{m-1} {{x}^2_{N-1-n}}\\
	{S^{(c)}_\tau} = \sum\limits_{n=0}^{N-1-m} {{x}^2_{n+m }}
		= \sum\limits_{n=0}^{N-1} {{x}^2_{n}} - \sum\limits_{n=0}^{m-1} {{x}^2_{n}}.
	\end{gather}
\end{subequations}
These need to be combined and rearranged into a form suitable for implementation in computer code
\begin{equation} \label{eq:SFfast}
	\begin{split}
	{S_\tau} = \frac{1}{(N-m)} \left(
	2N\left( R^\prime_0(x_n)-R^\prime_\tau (x_n)\right) 
	- \sum\limits_{\substack{ n=0 \\ m>0 }}^{m-1} \left( {{x}^2_{N-1-n}} + {{x}^2_{n}}\right)
	 \right)
	\end{split}
\end{equation}
where, as above, $\tau = m \Delta t$ for $m = 0 \dotsc (N-1)$. The CPU time required to evaluate
the SF in this form when $N$ is an integer power of two is $O(N\log N)$, see Table~\ref{tab:CPU}.
For clarity, this method and its results will be referred to as the \textit{fast structure function} (FSF). 

\begin{table}
	\centering
	\caption{Comparison of the CPU (3.06\,GHz Intel Core i3 processor) time needed to evaluate the SF
	using the direct (eqn~\ref{eq:SF}) and fast (eqn~\ref{eq:SFfast}) methods for a range of sample sizes, $N$.}
	\label{tab:CPU}
	\begin{tabular}{rccc} % four columns, alignment for each
		\hline
		 Samples & $N$ & {Direct (s)} & {Fast (s)} \\
		\hline
		\rule{0pt}{1.2em}
		4\,K &   $2^{12}$   &  \phantom{000}0.013                   &  \phantom{00}0.001 \\
		32\,K & $2^{15}$   & \phantom{000}0.61\phantom{0}   &  \phantom{00}0.008 \\
		256\,K & $2^{18}$ &  \phantom{00}38.2\phantom{00}  &  \phantom{00}0.072 \\
		2\,M &  $2^{21}$   &  4090\phantom{.000}                    & \phantom{00}0.84\phantom{0} \\
		16\,M & $2^{24}$  &                                                      & \phantom{00}6.8\phantom{00} \\			   
		32\,M & $2^{25}$  &                                                      & \phantom{0}13.7\phantom{00} \\			   
		64\,M & $2^{26}$  &                                                      & \phantom{0}29\phantom{.000} \\			   
		128\,M & $2^{27}$ &                                                     & 116\phantom{.000} \\			   
		\hline
	\end{tabular}
\end{table}

\subsection{Practicalities}\label{sec:realisationExamples}

The demonstration code in the supplementary material gives details of the FSF implementation.
Power-law noise was created in the following way.
We used the Box-Muller transform \citep{box1958} to realise Gaussian variates, $\mathcal{N}(0,\sigma^2)$.
Distributions with a specified cumulative distribution function (CDF), $\mathcal{F}(X)$, are realised using the inversion method \citep[][\S{III.2}]{devroye1986}, \textit{i.e.} solving $\mathcal{F}(x) = U[0,1)$, where $U[0,1)$ is a uniform variate we generated using the Mersenne Twister algorithm SFMT19937 \citep{Saito2008}. 
We then solved for $\mathcal{F}(x)$ numerically using a binary search, accelerated by guide tables, followed by interpolation.

\section{The photon and instrument noise}
\label{app:bg}

\subsection{Modelling the photon and instrument noise}
\label{sec:mod_bg}

\begin{figure}
\begin{center}
\includegraphics[width=\columnwidth]{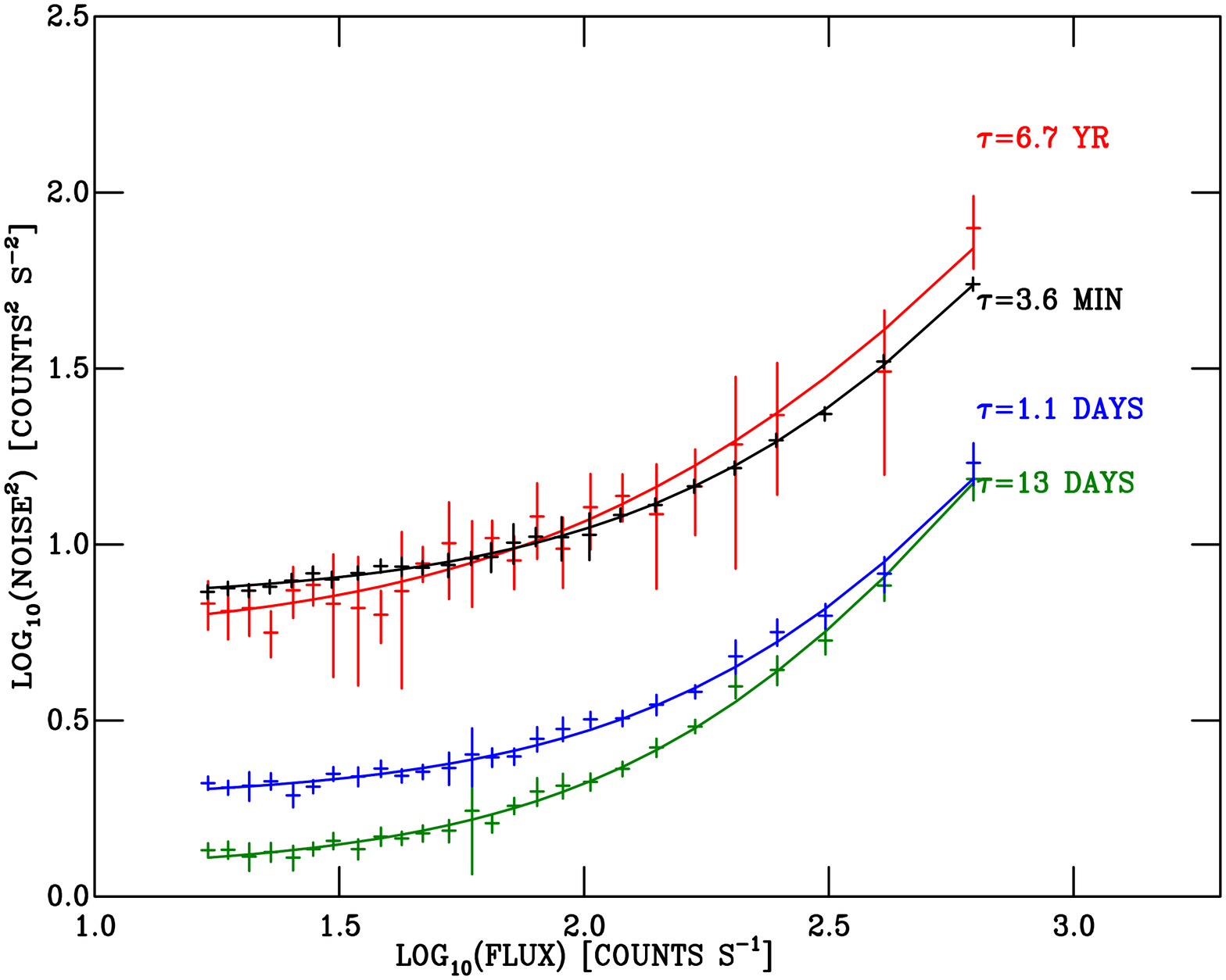}\\
\caption{Example noise model for the local photometric standard stars on CCD 4. 
Each point represents a slice of 51 stars in flux, in a given SF timescale bin. 
The data are the measured values, the lines are the best-fit analytical model described by Equation \ref{eq:noise_mod}. 
The error bars represent the RMS of the measured values.
The different colours represent different timescales.}
\label{fig:noise_model}
\end{center}
\end{figure}

As we discussed in Section \ref{sec:noise}, it is desirable to remove the photon and instrument noise from the SF, using the local photometric standard stars to model that noise as a function of timescale.
Of course the noise is also a function of the magnitude of the star, and so we need to use LPSs close in brightness to our targets.
\cite{Kozowski:2010aa} achieve this by using the SFs of four stars with similar magnitudes to the target.
In contrast we modelled the noise as a function of magnitude using a semi-emipircal model, and then interpolated the models for the magnitudes of our targets.
To achieve this the SFs for the local photometric standard stars on each CCD were ordered by mean flux and then sliced into groups of 51 stars. 
The median SF as a function of $\tau$ was calculated for each 51-star magnitude slice.
The SF at a particular timescale was then extracted from each of these median SFs to 
obtain the empirical SF as a function of flux at each timescale (see Fig. \ref{fig:noise_model}).
Next, for each timescale the SF was converted into flux and fitted as a function of flux with an analytical model of the form
\begin{equation}\label{eq:noise_mod}
F^2S=\epsilon = { B^{2} + K_{1}F+ K_{2}F^{2}}.
\end{equation}
Here $B$ is the noise due to the sky background in the frame (a fitted parameter that is constant across the CCD), $F$ is the measured stellar flux and $K_{1}$ and $K_{2}$ are empirically derived constants. 
This formula is motivated in the following way.
The first term represents the idea that for very faint stars the noise in their flux is dominated by the photon noise counts from the sky, and so the noise is independent of the flux \citep[see, for example][]{1998MNRAS.296..339N}.
Once the star becomes significantly brighter than the sky, the photon noise from the star itself dominates, giving the second term where the noise is proportional to the square root of the flux (which is then squared to be added in quadrature to the sky term).
Finally, for the brightest stars errors in the flat field will dominate, giving a term which varies linearly with stellar flux.
However, the ultimate justification for the form of Equation \ref{eq:noise_mod} is that the model fits the data (especially as roughly 10 percent of models have negative, i.e. unphysical values of $B$ or $K_1$).
Examples of the empirical relations and analytical fits are shown in Fig. \ref{fig:noise_model}.

\subsection{Subtracting the photon and instrument noise}
\label{sec:subtraction}

These models characterise the SF of the photon and instrument noise, and can be interpolated to the flux of each target star for a given timescale, CCD and stellar flux. 
This was then subtracted from each target SF using the formula which assumes the corresponding values of $Y$ are uncorrelated.
\begin{equation}
S_{\rm nett}=S - S_{\rm LPS}
\end{equation}
A small number of objects were so bright that they exceeded the highest flux datapoint in the fit.
In these cases we used the noise for the brightest fitted value rather than extrapolating, which gives an over-estimate of the fractional noise.

\section{Uncertainties for structure functions}
\label{app:uncer}

\subsection{A model for structure function uncertainties}

Our definition of the SF (Equation \ref{eq:sf}) can be interpreted as averaging many individual estimates of its value, in which case one might expect the RMS of the individual values about that mean to give an estimate of the uncertainty in an individual value.
This could then be divided by the square root of the number of contributing datapoints ($p^{1/2}$) to yield a standard error, which is an estimate of the uncertainty in the mean.
This intuition is almost correct.

For brevity of notation let us begin by defining $Y_{ij}$ as $(F_i-F_j)/F_{\rm med}$.
If our values of $Y$ are distributed as a Gaussian about a mean value of zero, then we expect $Y^2$ to be distributed as $\chi^2$ with one degree of freedom, scaled by the $\sigma^2$ of the Gaussian distribution characterising $Y$. 
The SF is the mean of this distribution, and the RMS does provide a measure of its width (albeit that its distribution is not Gaussian). 
Hence we could estimate the uncertainty in the mean as 
\begin{equation}
\Delta S = {1\over p_{\rm eff}^{1/2}}\sqrt {{1\over p} \sum_{ij}({ Y_{ij}^2} -S)^2}
\label{eq:whats_n_eff}
\end{equation}
where the following discussion centres around the value of $p_{\rm eff}$, which is the effective number of independent datapoints.
The reason $p_{\rm eff}$ is not one is that a given datapoint can contribute to many values of $Y$, and so the values of $Y$ are not independent.
Without making the above arguments explicit, \cite{2001ApJ...555..775C} suggest $p_{\rm eff}=p/2$, which they acknowledge is ad hoc, but gives reasonable estimates of the uncertainties.
Conversely, re-arranging Equation B1 of \cite{Czerny:2003aa} gives $p_{\rm eff}=p/4$.
Part of the issue here may be the non-Gaussian nature of the distribution, but an equally serious issue is the number of independent data points.
Were the SF derived from 2$p$ lightcurve points each of which was only paired with one other data point to create $p$ values of $y$ then the RMS should clearly be divided by $p^{1/2}$.
But in constructing a value of the SF many lightcurve points are used many times, and so the number of truly independent points is probably the number of lightcurve points used.
We confirmed this by creating a single SF value using every possible pairing of a set of $\nu$ values drawn from a Gaussian distribution.
We then used the RMS of the values of $Y_{ij}^2-S$ divided by $\nu^{1/2}$ to estimate the uncertainty for this SF value.
If we then created many SF values in this way we found their RMS matched the predicted uncertainty, apparently confirming $p_{\rm eff}$ should be the number of lightcurve points used.

\begin{figure}
\begin{center}
\includegraphics[width=\columnwidth]{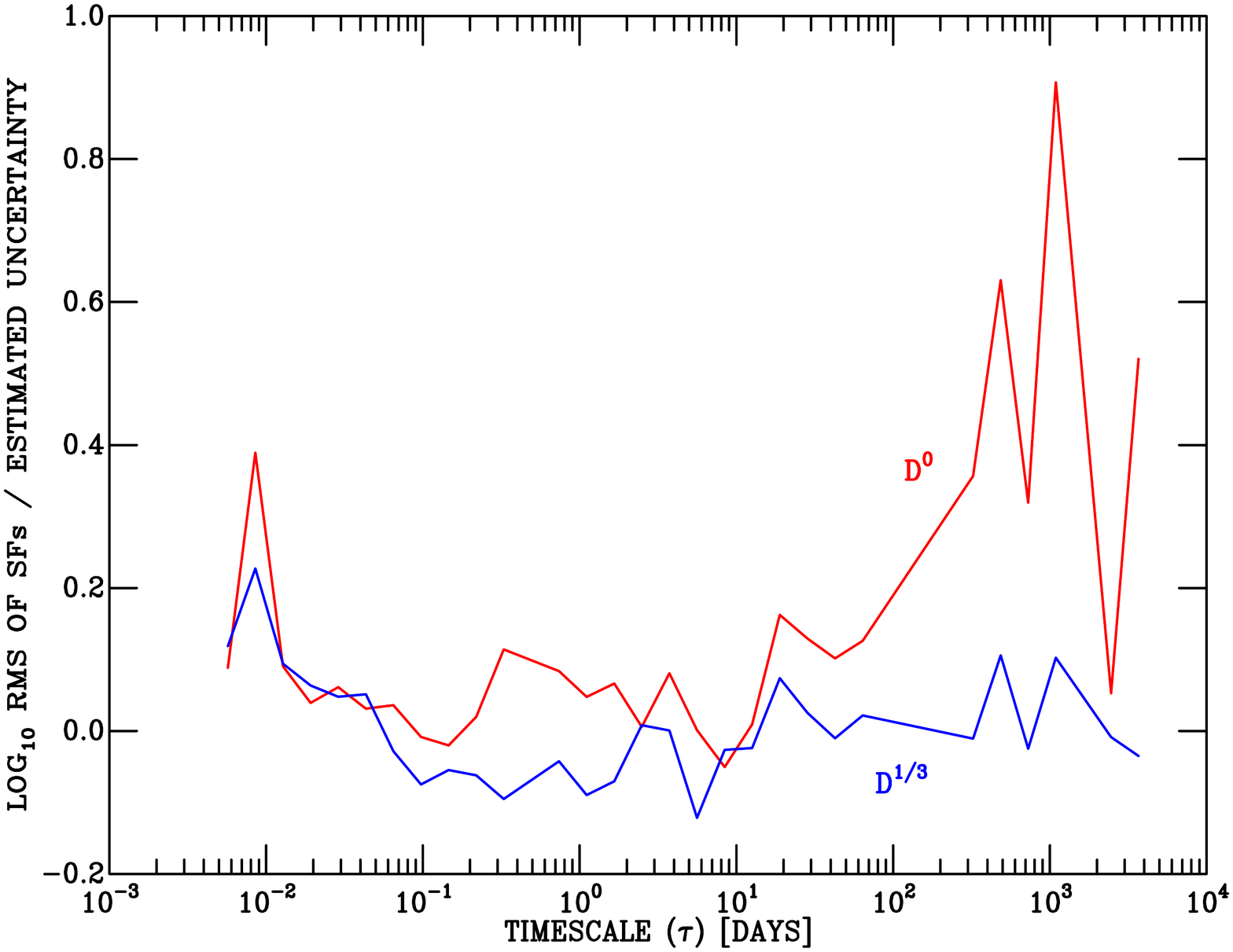}\\
\caption{
The RMS between sets of simulated SFs compared with the estimate of the noise given by Equation \ref{eq:whats_n_eff}.
The simulated SFs used the same time sampling as the low-cadence data, and points drawn from a Gaussian distribution.
The red (upper) line sets the effective number of datapoints ($p_{\rm eff}$) to be the number of datapoints in the lightcurve ($\nu$), whereas the lower (blue) curve uses $p_{\rm eff} = \nu/D^{1/3}$.
}
\label{fig:test_n_eff}
\end{center}
\end{figure}

Unfortunately in real SFs a single lightcurve point may be used in many values of $Y$, or just one, meaning that a small number of lighcurve points can dominate the value of the SF at a particular timescale, so the number of independent points is effectively reduced.
To illustrate this we created simulated lightcurves on the sampling of our low-cadence data, where each lightcurve point was drawn from a Gaussian distribution.
We then compared the RMS values of the SFs at each timescale with the RMS predicted by
Equation \ref{eq:whats_n_eff} if $p_{\rm eff}$ is set to one.
As Fig. \ref{fig:test_n_eff} shows this is a good way to estimate the uncertainties in the SF, giving answers which are within about a factor two of the true answer for timescales less than a few hundred days.
The timescales where this does not work are indeed dominated by a few datapoints.
(This is the same effect as we illustrated in Figure \ref{fig:walk_sfs}.)
One measure of this dominance is to examine distribution of the number of times each lightcurve point is used in creating a given point of the SF, and take the maximum of this distribution divided by its median. 
We will call this metric $D$.
In principle we could simulate a correction factor for each SF using the number of times each lightcurve point is used at each timescale, which would vary (due to flagged datapoints) from star to star.
In fact we found that if we use $p_{\rm eff} = \nu/D^{1/3}$ we have a good estimate of the noise (see Fig. \ref{eq:whats_n_eff}), though with no better justification for the power of $D$ than it works for our low-cadence data, and makes little difference for our high-cadence data.

\subsection{A test for structure function uncertainties}

\begin{figure}
\begin{center}
\includegraphics[width=\columnwidth]{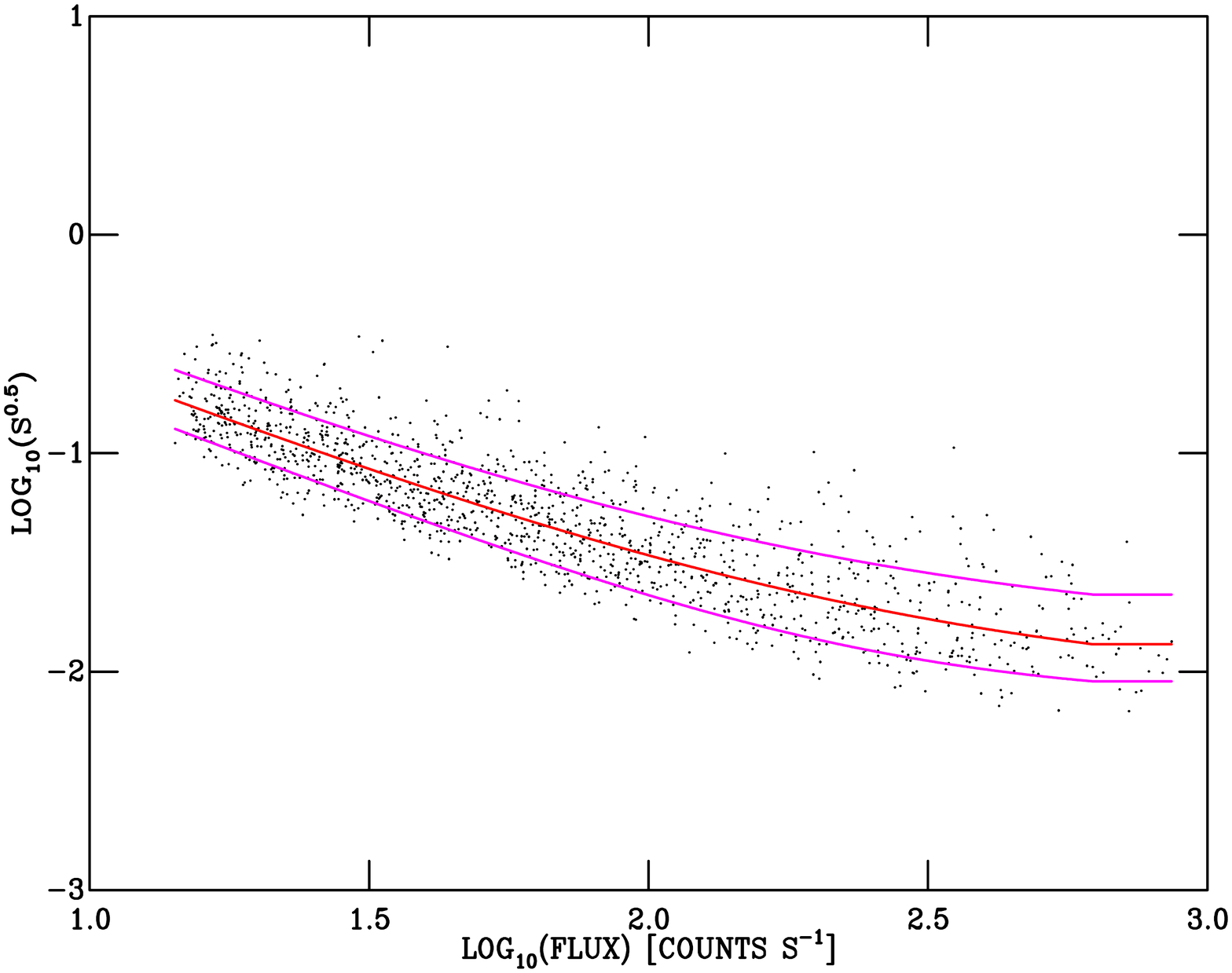}\\
\caption{The distribution of the values of the SFs at $\tau$=6.7 years for the local photometric standard stars on CCD 4.
Each point is the value of the SF plotted at the median flux for that star.
The red line was constructed by first dividing the data by flux into groups of 51 stars.
A polynomial was then fitted to the median value of the SFs in each group, and so it represents the 50th percentile by SF, and is identical to the red line in Fig. \ref{fig:noise_model}, though with the ordinate in SF rather than noise squared (i.e. different by a factor of the flux squared).
The fuchsia-coloured lines are the 16th and 84th percentiles calculated in the same way.
}
\label{fig:bg_uncer}
\end{center}
\end{figure}

We tested Equation \ref{eq:whats_n_eff} using the local photometric standards in a similar fashion to the way we used them to find the photon and instrument noise.
Fig. \ref{fig:bg_uncer} shows that there is scatter about the mean relationship for the photon and instrument noise.
That scatter represents the uncertainty in the measurement of the SF, and so in the same way as we fitted the median value in Section \ref{sec:mod_bg}, we can fit the 16 and 84 percentage points at each value of the flux
(given in electronic Tables C and D), as shown in Fig. \ref{fig:bg_uncer}.
In principle these are equivalent to the $\pm$1$\sigma$ limits enclosing 68 percent of the SFs, but of course the distribution is distinctly non-Gaussian.
For example the 16th percentile point is closer to the median line than the 84th percentile point by, on average a factor of $\simeq$0.7.

\begin{figure}
\begin{center}
\includegraphics[width=\columnwidth,angle=270]{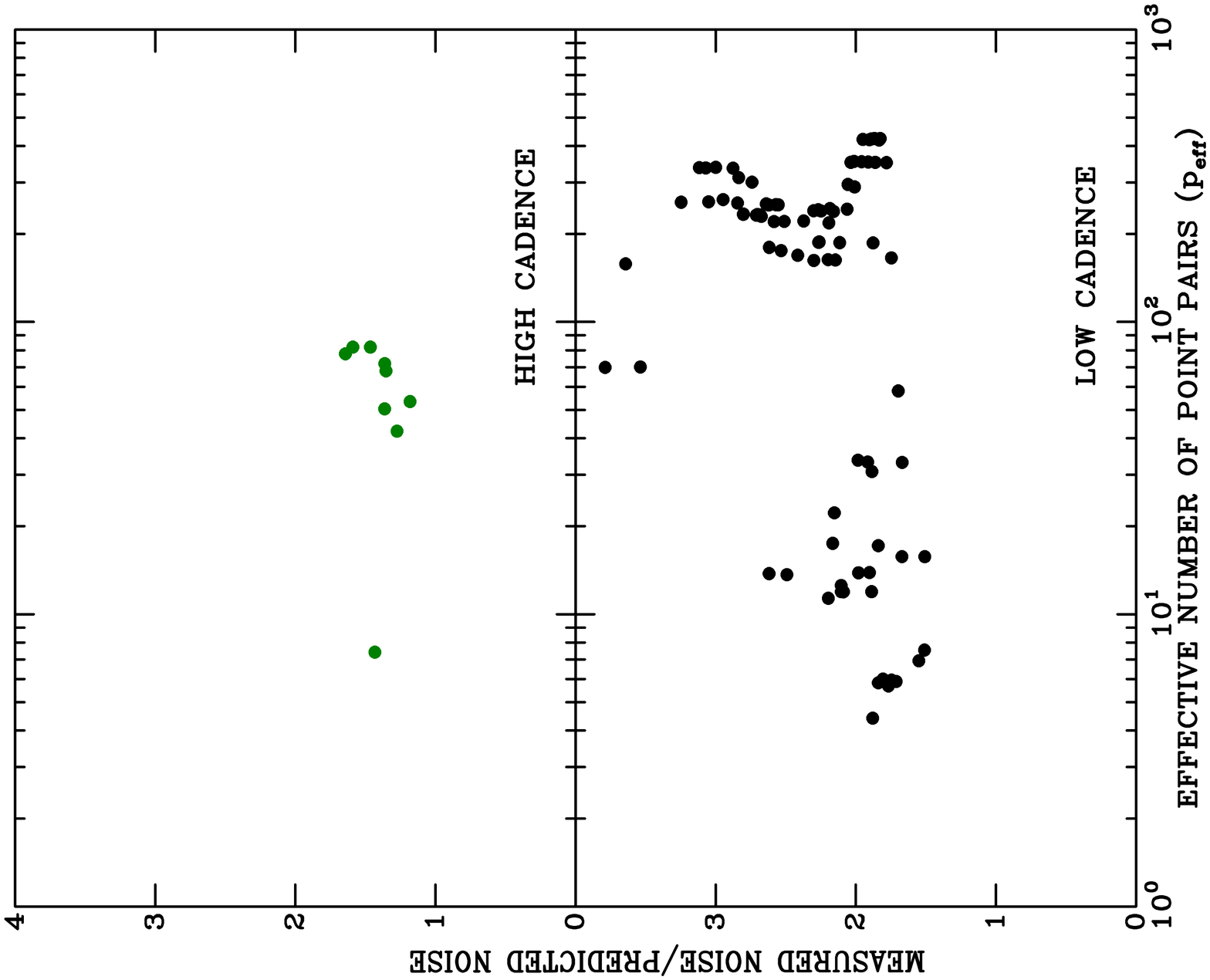}\\
\caption{
The median ratio of measured noise to estimated uncertainty as a function of the effective number of datapoints.
}
\label{fig:uncer_med}
\end{center}
\end{figure}

For each LPS we compared the ratio of the model noise for every SF timescale as given Equation \ref{eq:whats_n_eff}, to the measured differences in the SFs for an ensemble of stars of similar brightness, given by the half the difference between the 16th and 84th percentile points.
We then collected together all the resulting datapoints which had the same effective number of datapoints in the LPS (though different mean fluxes), calculated the median ratio and plotted this as a function of the number of point pairs.
This is shown in Fig. \ref{fig:uncer_med}.
It is clear than in both the low and high cadence datasets we are still under-estimating the uncertainties, by a factors of 2.3 and 1.4 respectively.
There is also a weak residual dependence with $p_{\rm eff}$ for the low cadence data.

\begin{figure}
\begin{center}
\includegraphics[width=\columnwidth,angle=270]{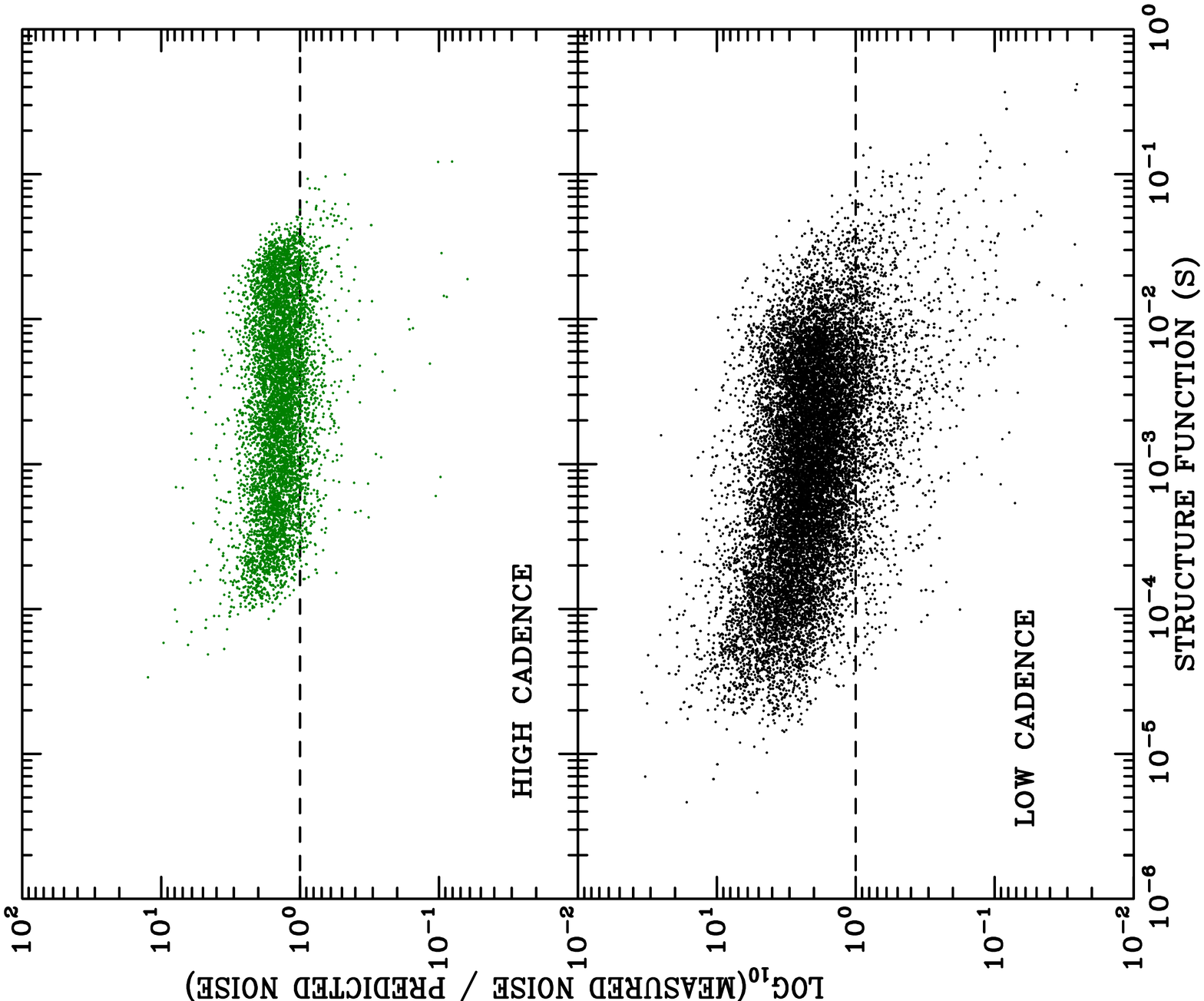}\\
\caption{The ratio of measured noise to estimated uncertainty as a function for one in ten of the measured values of the SFs of the LPS.
}
\label{fig:uncer_all}
\end{center}
\end{figure}

We then examined whether there was any residual dependence with $S$ (the SF before it is corrected for photon and instrument noise). 
In this case, though, rather than plotting the median for all points with the same number of point pairs, in Fig. \ref{fig:uncer_all} we take a sample from the entire dataset of SF values for the LPS and plot them as a function of $S$.
In this figure both datasets show the same offsets as the plots as a function of $p_{\rm eff}$, with the ratio of measured to the predicted noise is constant over most of the range of $S$.
Only below an RMS of $\sim$0.01 ($S$$\sim$$10^{-4}$) is there a significant deviation from this.

\subsection{The final model uncertainties}

The main issue to be addressed, therefore, is the global offset roughly independent of $S$ and $p_{\rm eff}$.
The underlying physical reason for this difference between random simulated data and real data is unclear, but it is clear we must correct for it.
Therefore to evaluate the uncertainties in $S_{\rm nett}$ we use Equation \ref{eq:whats_n_eff} with $p_{\rm eff}$=$\nu/D^{1/3}$, and apply the multiplicative factor for the low and high cadence datasets given above.

%%%%%%%%%%%%%%%%%%%%%%%%%%%%%%%%%%%%%%%%%%%%%%%%%%

% Don't change these lines
\bsp	% typesetting comment
\label{lastpage}
\end{document}